\documentclass[aps,pra,twocolumn,amsfonts,amssymb,amsmath,showpacs,showkeys,
floatfix,nofootinbib,groupedaddress,superscriptaddress,citesort]{revtex4}%
\usepackage{mathrsfs}
\usepackage{amsfonts}
\usepackage{amstext}
\usepackage{amsmath}
\usepackage{amssymb}
\usepackage{bm}
\usepackage{bbm}
\usepackage{graphicx}
\def\qed{\leavevmode\unskip\penalty9999 \hbox{}\nobreak\hfill
     \quad\hbox{\leavevmode  \hbox to.77778em{%
              \hfil\vrule   \vbox to.675em%
               {\hrule width.6em\vfil\hrule}\vrule\hfil}}
     \par\vskip3pt}

\newtheorem{theorem}{Theorem}

\newtheorem{definition}{Definition}
\begin{document}

\preprint{APS/123-QED}

\title[A  multipartite multimode Gaussian quantum correlation]
{A computable multipartite multimode Gaussian quantum correlation measure
and the monogamy relations for  continuous-variable systems}

\author{Jinchuan Hou}\thanks{Corresponding author; jinchuanhou@aliyun.com}
\affiliation{College of Mathematics, Taiyuan University of
Technology, Taiyuan 030024, P. R. China}

\author{Liang Liu}
\affiliation{College of Mathematics, Taiyuan University of
Technology, Taiyuan 030024, P. R.
China}\email{liuliang@tyut.edu.cn}

\author{Xiaofei Qi}\thanks{Corresponding author; xiaofeiqisxu@aliyun.com}
\affiliation{School of Mathematical Science, Shanxi University,
Taiyuan 030006, P. R. China}

\begin{abstract}
In this paper, definitions of the unification condition, the hierarchy condition
and  three kinds of monogamy relations for  multipartite quantum
correlation measures are given and discussed.
 A computable multipartite multimode Gaussian quantum correlation
measure ${\mathcal M}^{(k)}$ is proposed for any $k$-partite
multimode continuous-variable systems with $k\geq 2$.
The value of ${\mathcal M}^{(k)}$ only depends  on the
covariance matrices of  continuous-variable states, is invariant under any permutation
of subsystems, has no ancilla problem, is nonincreasing under
$k$-partite local Gaussian channels (particularly, invariant under
$k$-partite locally Gaussian unitary operations), and vanishes on
$k$-partite product states. For a $k$-partite Gaussian state $\rho$,
${\mathcal M}^{(k)}(\rho)=0$ if and only if $\rho$ is a $k$-partite
product state.  Moreover,
 ${\mathcal M}^{(k)}$ satisfies the unification condition and the hierarchy
 condition that a multipartite quantum correlation measure should obey.
We also show that ${\mathcal M}^{(k)}$ is not strongly monogamous,
but completely monogamous and tightly
monogamous.
\end{abstract}


\maketitle

\section{Introduction}

An amazing feature of quantum mechanics is the presence of quantum correlations in composite quantum systems.   It is
proved that quantum correlations beyond entanglement can
also be exploited in quantum information tasks as physical
resources.  Various methods have been proposed to describe
bipartite quantum correlations, such as quantum discord
\cite{Ollivier}, geometric quantum discord
\cite{Borivoje,Luo,Miranowicz}, measurement-induced nonlocality
(MIN) \cite{Luo-Fu} and measurement-induced disturbance (MID)
\cite{Luo-S} for discrete-variable systems.  Notice that, in many
quantum protocols, the systems considered are continuous-variable
(CV) systems.  Therefore, it is also  important and
interesting to study quantum correlations in CV
systems.

Denote by ${\mathcal {GS}}^{m+n}(H_A\otimes H_B)$ the set of all
$(m+n)$-mode Gaussian states in the CV system
described by a Hilbert space $H_A\otimes H_B$. Let ${\mathcal
G}_{A/B}: {\mathcal {GS}}^{m+n}\to [0,+\infty)$ be a functional.
Following the idea from
\cite{ModiBPV,Ciccaarello,Filgueiras,RogaIll}, ${\mathcal G}_{A/B}$
is a Gaussian quantum correlation measure (GQCM) for a Gaussian
quantum correlation (GQC) with respect to subsystem A/B if it
satisfies the following four conditions:

i) for any Gaussian state $\rho_{AB}$, ${\mathcal G}_{A/B}(\rho_{AB})=0$ if and only if $\rho_{AB}$ contains no  GQC;

ii) (Locally Gaussian unitary invariant) ${\mathcal G}_{A/B}((W\otimes
    V)\rho_{AB}(W^{\dag}\otimes V^{\dag}))={\mathcal G}_{A/B}(\rho_{AB})$ holds for any Gaussian unitary operators $W$ on $H_A$, $V$ on $H_B$ and any Gaussian state $\rho_{AB}$;

iii) (Non-increasing under local Gaussian channels performed on B/A) ${\mathcal G}_{A}((I\otimes
    \Phi_B)\rho_{AB})\leq{\mathcal G}_{A}(\rho_{AB})$ (resp. ${\mathcal G}_{B}((\Phi_A \otimes I)\rho_{AB})\leq{\mathcal G}_{B}(\rho_{AB})$) holds for any Gaussian channel $\Phi_B$ (resp. $\Phi_A$)
    performed on subsystem B (resp. on subsystem A) and any Guassian state $\rho_{AB}$;

iv)  ${\mathcal G}_{A/B}$ describes the entanglement on pure
Gaussian states; that is, if $|\psi\rangle\langle\psi|$ is a pure
Gaussian state, then ${\mathcal
G}_{A/B}(|\psi\rangle\langle\psi|)=0$ if and only if $|\psi\rangle$
is a product state.\\
Furthermore,
${\mathcal G}_{A}$ is a  nice GQCM if it satisfies i)--iii)
and the following:

iv$'$) (Reducing to an entanglement measure for pure states) There
exists an entanglement monotone $\mathcal E$ such that ${\mathcal
G}_{A}(|\psi\rangle\langle\psi|)={\mathcal
E}(|\psi\rangle\langle\psi|)$ for any bipartite Gaussian pure state
$|\psi\rangle\langle\psi|$.

Several GQCMs have been proposed for bipartite CV
systems. Giorda, Paris \cite{Giorda} and Adesso, Datta \cite{Adesso}
independently gave the definition of Gaussian quantum discord $D$
for Gaussian states.   Adesso and  Girolami in
\cite{Adesso-Girolami} proposed the concept of Gaussian geometric
discord $D_G$. It was shown that, for a Gaussian state $\rho_{AB}$,
$D(\rho_{AB})=0$ ($D_G(\rho_{AB})=0$) if and only if $\rho_{AB}$ is
a product state; that is, $\rho_{AB}$ has no quantum correlation if
and only if it is a product state. After then, remarkable efforts have
been made to find simpler ways to quantify this Gaussian correlation. For instance,
MID of Gaussian states was proposed \cite{Mista} and MINs
$Q,Q_{\mathcal P}$ for Gaussian states was studied \cite{Ma}.
Gaussian discriminating strength
 based on the minimum or maximum change induced on the
 state by  a locally Gaussian
unitary operation were investigated in \cite{Farace, Rigovacca, WHQ}.
Based on Gaussian unitary operation and the fidelity,  several kinds
of Gaussian response of discord (for example,
$\mathcal{G}D_{R}^{x}$, $N_F$, $N_{\mathcal F}$) were proposed and
discussed in \cite{Roga-Illuminati, Liu}. For other related
results, see \cite{MHQ, Fu, Datta, Gharibian,LHAW} and the references
therein. All
 quantifications mentioned above describe the same GQC as that described
 by Gaussian quantum discord.

 However,  no one of the bipartite GQCMs mentioned above is
easily  accessible.  It is very difficult to calculate the values for all $(n+m)$-mode Gaussian states except
$(1+1)$-mode Gaussian states or some special Gaussian states since
these GQCMs involve some
measurements on a subsystem and some optimization process. Also note that, these
GQCMs are not symmetric about the subsystems though
the corresponding GQCs are. The second point is that, these
GQCMs can not be extended to multipartite systems
evidently.

Thus, two problems  arise.

{\bf Problem 1}. Whether or not there exist  some ways of
quantifying GQCs for bipartite CV systems  that are easily accessible?

{\bf Problem 2}.
What are the rules that  every multipartite multimode GQCM (beyond entanglement)
should obey and whether or not there exist such  GQCMs for multipartite CV systems that are easily accessible, and furthermore, are  monogamous in some sense?

For the first problem, an effort is made in \cite{LHQ}, where a computable GQCM $\mathcal
M$ for $(n+m)$-mode CV systems is proposed. It is shown that $\mathcal M$ satisfies the
following nicer properties:

(1)  for any $(n+m)$-mode Gaussian state $\rho_{AB}$, $\mathcal
M(\rho_{AB})=0$ if and only if $\rho_{AB}$ is a product state;

(2) $\mathcal M$ is locally Gaussian unitary invariant;

(3) $\mathcal M$ is non-increasing under local  Gaussian channels in
the sense that $\mathcal M ((\Phi_A\otimes \Phi_B)\rho_{AB})\leq
\mathcal M(\rho_{AB})$ holds for any Gaussian channel $\Phi_{A/B}$
performed on subsystem A/B and any Gaussian state $\rho_{AB}$;

(4) $\mathcal M$ is independent of the mean, is symmetric about
subsystems  and has no ancilla problem.

But, for the second problem, by our knowledge, no results of quantifying GQCs (beyond entanglement) for multipartite multimode CV systems were known.
The purpose of this paper is to give an answer to the second problem.

Not like the bipartite GQCM, as a multipartite multimode GQCM, it should obey some additional  rules.
  For   multipartite entanglement
measures, these additional rules were discussed firstly in \cite{GuoZhang}.   It is pointed in  \cite{GuoZhang} that a multipartite entanglement
measure should meet the unification condition and the hierarchy
condition. For a quantum correlation beyond entanglement, as a physical resource, it is reasonable to require that the unification
condition and the hierarchy condition should also be obeyed by their
multipartite quantum correlation measures. The unification condition is easily understood, but the hierarchy condition is not defined clearly in \cite{GuoZhang}.

In this paper, we
give exactly a definition of the hierarchy condition which declares
that the whole correlation of lower partition is not greater than
the whole correlation of higher partition;  the partial correlation
is not greater than the whole correlation;  and the correlation
after kicking some parties out of subgroups is not greater than the
correlation between the subgroups.
We also propose  a
multipartite multimode GQCM  ${\mathcal M}^{(k)}$  and discuss its
properties for any $k$-partite  CV  systems
($k\geq 2$). The definition of ${\mathcal M}^{(k)}$ only depends on
the covariance matrix of  CV states and thus is more easily
calculated  for any CV state with finite second moments. ${\mathcal
M}^{(k)}$ is a multipartite extension  of $\mathcal M$ as $\mathcal
M^{(2)}=\mathcal M$. We show that ${\mathcal M}^{(k)}$ has almost
all expected good properties: ${\mathcal M}^{(k)}$ vanishes on
$k$-partite product states and, for a $k$-partite Gaussian state
$\rho$, ${\mathcal M}^{(k)}(\rho)=0$ if and only if $\rho$ is a
$k$-partite product state; ${\mathcal M}^{(k)}$ is invariant under
any permutation of subsystems; has no ancilla problem; is
nonincreasing under $k$-partite local Gaussian channels
(particularly, is invariant under $k$-partite locally Gaussian unitary
operations) in the sense that ${\mathcal M}^{(k)}((\Phi_1\otimes
\Phi_2\otimes \cdots\otimes \Phi_k)\rho_{A_1,A_2,\ldots, A_k})\leq
{\mathcal M}^{(k)}(\rho_{A_1,A_2,\ldots ,A_k})$ for any Gaussian
channel $\Phi_j$ in subsystem $H_{A_j}$ and any Gaussian state
$\rho_{A_1,A_2,\ldots, A_k}$.
More importantly, we show that ${\mathcal M}^{(k)}$
satisfies the unification condition and the hierarchy
 condition that a multipartite quantum correlation measure should obey. Therefore, it is true that ${\mathcal M}^{(k)}$ is an accessible
multipartite multimode Gaussian quantum correlation measure for CV systems.

 Finally, the monogamy relation for ${\mathcal M}^{(k)}$ is
 investigated.

 Recall that a bipartite entanglement measure
$E$ is monogamous if $E(\rho_{A|BC})\geq E(\rho_{AB})+E(\rho_{AC}) $ (CKW inequality)
holds for any $\rho_{ABC}$ \cite{CKW}.  Many bipartite entanglement measures  are monogamous (see Ref. \cite{KW, OV, GG} and the references therein).
In \cite{GG},
Gour and Guo proposed the monogamy without inequalities.
It seems that the monogamy relation (i.e., CKW inequality) is a natural feature for quantum entanglement, because
entanglement is a kind of physical resource and thus the amount  of part entanglements cannot exceed the amount of total entanglement, which is almost equivalent to the statement that if
two parties A and B are maximally entangled, then neither of
them can share entanglement with a third party C \cite{GuoZhang, GG}.
The monogamy relations
for quantum correlations beyond entanglement have also been
investigated \cite{GLGiorgi,Streltsov}. But it is surprising that all bipartite quantum correlation measures beyond entanglement for discrete systems, including quantum discord, are not monogamous in general \cite{Streltsov},
which is a contradiction to the fact that many quantum correlations beyond entanglement are physical resources.
The trouble may come from the definition of monogamy relations. The monogamy relation discusses relationships between three parties by using a bipartite measure, focuses only on the relation between the parties A, BC, the
parties A, B as well as the parties A, C, and ignores the relation
contained in the parties ABC and the relation between parties B, C,
which seems incomplete. In fact, by the hierarchy
 condition,   $E(\rho_{A|BC})$ is still a part of the total entanglement  $E^{(3)}(\rho_{ABC})$
shared by A, B and C, where $E^{(3)}$ is a tripartite entangled measure and is consistent with $E$.  So, to understand the monogamy relation better, one should consider the question
in framework of multipartite entanglement measures. Guo and Zhang gave a strict framework for
defining multipartite entanglement measures, based on which,
the complete monogamy relation and the tight monogamy
relation were established \cite{GuoZhang}.

We give exactly the definitions of the monogamy relations for a multipartite quantum correlation measure in this paper. It is revealed that there
are three kinds of monogamy relations: (1) the tight monogamy
relation, which claims that the correlation between subgroups
attains the total correlation will imply that the parties in the
same subgroup are not correlated to each other; (2) the complete
 monogamy relation, which claims that the correlation of a
subgroup attains the total correlation will imply that the parties
out of the subgroup are not correlated with any other parties of the
system; and (3) the strong monogamy relation, which claims that the
correlation between subgroups after ``kicking some parties out of''
each subgroups keeps invariant will imply that the remain parties
are not correlated with the parties kicked out of. Then, we prove that
${\mathcal M}^{(k)}$ is completely monogamous and tightly
monogamous. However, ${\mathcal M}^{(k)}$ is not strongly monogamous.

 The paper is organized as follows. In
 Section 2, we recall briefly some notions and notations from
CV systems and propose the quantity $\mathcal M^{(k)}$.
 Section 3 is devoted to studying the basic properties of $\mathcal M^{(k)}$. In Section 4, we show that  ${\mathcal M}^{(k)}$
 satisfies the unification condition and the hierarchy
 condition. The monogamy relations for ${\mathcal M}^{(k)}$ are studied in Section 5.
 Finally,  a short conclusion is given in Section 6.

\section{Definition of $\mathcal M^{(k)}$}

Before giving our definition of the quantity $\mathcal M^{(k)}$, we need recall briefly some notions and notations concerning
Gaussian states  (for more details, ref. \cite{Sera}).

Recall that an $n$-mode continuous-variable system (CV system) is a
system  determined by $2n$-tuple
$(\hat{Q}_1,\hat{P}_1,\cdots,\hat{Q}_n,\hat{P}_n)$ of self-adjoint
operators with state space $H=H_1\otimes H_2\otimes\cdots\otimes
H_n$, where $\hat{P}_r,\hat{Q}_r$ are respectively
 the position and momentum operators of the $r$th-mode which act on the separable infinite
dimensional complex Hilbert space $H_r$. As it is well known,
$\hat{Q_r}=(\hat{a_r}+\hat{a_r}^\dag)/\sqrt{2}$ and
$\hat{P_r}=-i(\hat{a_r}-\hat{a_r}^\dag)/\sqrt{2}$
($r=1,2,\cdots,n$) with
 $\hat{a}_r^\dag$ and
$\hat{a}_r$ being the creation and annihilation operators in the $r$th
mode $H_r$, which satisfy the Canonical Commutation Relation (CCR)
$$[\hat{a}_r,\hat{a}_s^\dag]=\delta_{rs}I\ {\rm and}
\ [\hat{a}_r^\dag,\hat{a}_s^\dag]=[\hat{a}_r,\hat{a}_s]=0,\ \
r,s=1,2,\cdots,n.$$ Denote by ${\mathcal S}(H)$ the set of all
quantum states in a system described by $H$ (the positive operators on
$H$ with trace 1). The
characteristic function $\chi_{\rho}$ for  any state $\rho\in{\mathcal S}(H)$ is defined as
$$\chi_{\rho}(z)={\rm tr}(\rho W(z)),$$ where
$z=(x_{1}, y_{1}, \cdots, x_{n}, y_{n})^{\rm T}\in{\mathbb R}^{2n}$,
$ W(z)=\exp(i{R}z)$ is the Weyl displacement operator, ${R}=(\hat
R_1,\hat R_2,\cdots,\hat
R_{2n})=(\hat{Q}_1,\hat{P}_1,\cdots,\hat{Q}_n,\hat{P}_n)$.

Let ${\mathcal {FS}}(H)$ be the set of all quantum states with
finite second moments, that is, $\rho\in{\mathcal {FS}}(H)$ if ${\rm
Tr}(\rho \hat{R_r})<\infty$ and ${\rm Tr}(\rho \hat{R_r}^2)<\infty$
for all $r=1,2,\ldots, 2n$. For $\rho\in{\mathcal {FS}}(H)$, its
first moment vector
$$\begin{array}{rl}{\mathbf d}=&{\mathbf d}_\rho=(\langle\hat R_1 \rangle, \langle\hat R_2
\rangle, \ldots ,\langle\hat R_{2n} \rangle)^{\rm T}\\
=&({\rm tr}(\rho
\hat R_1), {\rm tr}(\rho \hat R_2), \ldots, {\rm tr}(\rho \hat
R_{2n}))^{\rm T}\in{\mathbb R}^{2n}\end{array}$$ and the second moment matrix
$$\Gamma=\Gamma_\rho=(\gamma_{kl})\in M_{2n}(\mathbb R)$$
defined by $\gamma_{kl}={\rm tr}[\rho
(\Delta\hat{R}_k\Delta\hat{R}_l+\Delta\hat{R}_l\Delta\hat{R}_k)]$
with $\Delta\hat{R}_k=\hat{R}_k-\langle\hat{R}_k\rangle$
(\cite{Braunstein}) are called respectively the mean (or the
displacement vector) of $\rho$ and the covariance matrix (CM) of
$\rho$.  Here
$M_k(\mathbb R)$ stands for the algebra of all $k\times k$ matrices
over the real field $\mathbb R$. Note that a CM $\Gamma$ must be real symmetric and satisfy the
condition $\Gamma +i\Delta\geq 0$, where
$\Delta=\oplus_{r=1}^n\Delta_r$ with $\Delta_r
=\begin{pmatrix}0&1\\-1&0\end{pmatrix}$ for each $r$.
A Gaussian state $\rho\in{\mathcal {FS}}(H)$ is such a state of which the  characteristic function
$\chi_{\rho}(z)$ is of the form
\begin{eqnarray*}\chi_{\rho}(z)=\exp[-\frac{1}{4}z^{\rm T}\Gamma_\rho z+i{\mathbf
d}_\rho^{\rm T}z].
\end{eqnarray*}

A quantum channel (trace preserving
complete positive map) $\Phi$  is called a Gaussian channel if
$\Phi$ sends every Gaussian state into a Gaussian state. A unitary operator $U$ acting on $H$ is said to be Gaussian if  the unitary operation
$\rho\mapsto U\rho U^\dag$ is a Gaussian channel.

Let $\rho_{A_1,A_2,\ldots, A_k}\in{\mathcal {FS}}(H_{A_1}\otimes
H_{A_2}\otimes\cdots \otimes H_{A_k})$ be a CV state in a $k$-partite
$(n_1+n_2+\cdots +n_k)$-mode CV system. Then its CM can be
represented as
$$\Gamma_{\rho_{A_1,A_2,\ldots, A_k}}=\left(\begin{array}{cccc} A_{11} & A_{12} &\cdots & A_{1k}\\
A_{21} & A_{22} &\cdots & A_{2k} \\ \vdots & \vdots & \ddots &\vdots
\\ A_{k1} & A_{k2} & \cdots & A_{kk}  \end{array}\right),\eqno(1)
$$
where $A_{jj}\in M_{2n_j}(\mathbb R)$ is the CM of the reduced state
$\rho_{A_j}={\rm Tr}_{A_j^c}(\rho_{A_1,A_2,\ldots, A_k})$ of
$\rho_{A_1,A_2,\ldots, A_k}$,  $A_j^c=\{A_1,\ldots, A_{j-1},
A_{j+1},\ldots, A_k\}$, namely, $A_{jj}=\Gamma_{\rho_{A_j}}$, and
$A_{ij}=A_{ji}^T\in M_{2n_i,2n_j}(\mathbb R)$ for any $i,j\in\{
1,2,\ldots, k\}$ which reveals quantum correlation between
subsystems ${A_i}$ and $A_j$.

\begin{definition}
For any $(n_1+n_2+\cdots +n_k)$-mode $k$-partite state
$\rho_{A_1,A_2,\ldots, A_k}\in{\mathcal {FS}}(H_{A_1}\otimes
H_{A_2}\otimes\cdots \otimes H_{A_k})$, the quantity ${\mathcal
M}^{(k)}(\rho_{A_1,A_2,\ldots, A_k})$ is defined by
$${\mathcal M}^{(k)}(\rho_{A_1,A_2,\ldots, A_k})=
1-\frac{\det(\Gamma_{\rho_{A_1,A_2,\ldots, A_k}})}{\Pi_{j=1}^k \det(\Gamma_{\rho_{A_j}})},$$
where   $\Gamma_{\rho_{A_1,A_2,\ldots, A_k}}$ and
$\Gamma_{\rho_{A_j}}$ are respectively the covariance matrices of
$\rho_{A_1,A_2,\ldots, A_k}$ and $\rho_{A_j}$.
\end{definition}

Obviously, the function ${\mathcal M}^{(k)}: {\mathcal
{FS}}(H_{A_1}\otimes H_{A_2}\otimes\cdots \otimes H_{A_k})\to
[0,+\infty)$ satisfies $0\leq {\mathcal
M}^{(k)}(\rho_{A_1,A_2,\ldots, A_k})<1$ and is independent of the mean.
Particularly, for bipartite case, ${\mathcal M}^{(2)}$ is just the same as
$\mathcal M$ proposed in \cite{LHQ}.

\section{Basic properties of ${\mathcal M}^{(k)}$}

By Definition 1, it is clear that, for any $\rho_{A_1,A_2,\ldots,
A_k}\in{\mathcal {FS}}(H_{A_1}\otimes H_{A_2}\otimes\cdots \otimes
H_{A_k})$, the value ${\mathcal M}^{(k)}(\rho_{A_1,A_2,\ldots,
A_k})$ is easily calculated, avoiding performing any measurement
and any  optimization procedure. Furthermore, ${\mathcal M}^{(k)}$
has the following  properties.

1) ${\mathcal M}^{(k)}$ vanishes on product states.

2) ${\mathcal M}^{(k)}$ is invariant under any permutation of
subsystems, that is, for any permutation $\pi$ of $(1,2,\ldots, k)$,
denoting by $\rho_{A_{\pi(1)},A_{\pi(2)},\ldots, A_{\pi(k)}}$ the
state obtained from the state $\rho_{A_1,A_2,\ldots, A_k}$ by
changing the order of the subsystems according to the permutation
$\pi$, we have
$${\mathcal M}^{(k)}(\rho_{A_{\pi(1)},A_{\pi(2)},\ldots, A_{\pi(k)}})={\mathcal M}^{(k)}(\rho_{A_1,A_2,\ldots,
A_k}).
$$

3) ${\mathcal M}^{(k)}$ has no ancilla problem:
$$
{\mathcal M}^{(k)}(\rho_{A_1,A_2,\ldots,
A_k}\otimes\rho_C)={\mathcal M}^{(k)}(\rho_{A_1,A_2,\ldots, A_k})
$$
when considering the $k$-partition $A_1|A_2|\ldots |A_{k-1}|A_kC$ of
the $(k+1)$-partite system $A_1A_2\ldots A_kC$.

4) ${\mathcal M}^{(k)}$ is invariant under $k$-partite locally
Gaussian unitary operations.

5)  For any $(n_1+n_2+\cdots +n_k)$-mode $k$-partite  state
$\rho_{A_1,A_2,\ldots, A_k}\in{\mathcal {FS}}(H_{A_1}\otimes
H_{A_2}\otimes\cdots \otimes H_{A_k})$ with CM
$\Gamma=(A_{ij})_{k\times k}$ as in Eq.(1), $\mathcal
M^{(k)}(\rho_{A_1,A_2,\ldots, A_k})=0$ if and only if $A_{ij}=0$
whenever $i\not=j$. Particularly, if $\rho_{A_1,A_2,\ldots, A_k}$ is
a Gaussian state, then $\mathcal M^{(k)}(\rho_{A_1,A_2,\ldots,
A_k})=0$ if and only if $\rho_{A_1,A_2,\ldots, A_k}$ is a
$k$-partite product Gaussian state, that is, $\rho_{A_1,A_2,\ldots,
A_k}=\rho_{A_1}\otimes \rho_{A_2}\otimes\cdots \otimes \rho_{A_k}$.

6)  (Nonincreasing under local Gaussian channels)  For any Gaussian
state $\rho_{A_1,A_2,\ldots, A_k}\in{\mathcal {FS}}(H_{A_1}\otimes
H_{A_2}\otimes\cdots\otimes H_{A_k})$ and any local Gaussian channel
$\Phi_1\otimes \Phi_2\otimes\cdots\otimes\Phi_k$, we have
$$\mathcal M^{(k)}((\Phi_1\otimes \Phi_2\otimes\cdots\otimes\Phi_k)\rho_{A_1,A_2,\ldots, A_k})\leq \mathcal M^{(k)}(\rho_{A_1,A_2,\ldots, A_k}).$$
Particularly, ${\mathcal M}^{(k)}$ is locally Gaussian unitary
invariant.

Proofs of 4)-6) will be given in Appendix A.

 Thus, ${\mathcal M}^{(k)}$ ($k\geq 2$) is a possible
candidate of computable quantification of the multipartite multimode
GQC  for $k$-partite CV systems, which describes the natural quantum
correlation for Gaussian states that a state contains no correlation
if and only if it is a product state.

\section{Unification condition and Hierarchy condition for ${\mathcal M}^{(k)}$}

To show that ${\mathcal M}^{(k)}$ is a multipartite multimode
GQCM, we have to check further the
unification condition and the hierarchy condition for ${\mathcal
M}^{(k)}$.

The unification condition and the hierarchy condition were firstly
proposed in \cite{GuoZhang} for multipartite entanglement measure.
Recall that a bipartite entanglement measure $E$ is a nonnegative
functional on bipartite states which vanishes on separable states
and is nonincreasing under LOCC. However,  a multipartite
entanglement measure should satisfy some additional conditions
such as the {\it unification condition} and the {\it hierarchy
condition}. For example, for a tripartite entanglement measure
$E^{(3)}:{\mathcal S}(H_A\otimes H_B\otimes H_C)\to [0,+\infty)$,
apart from the usual requirements that $E^{(3)}$ vanishes on full
separable states and can not increase under 3-partite LOCC,
$E^{(3)}$ should satisfy further the unification condition (i.e.,
$E^{(3)}$ is invariant under the permutations of subsystems and
 a bipartite entanglement measure $E^{(2)}$ can be defined which is consistent with
$E^{(3)}$) and the hierarchy condition (i.e.,
$E^{(3)}(\rho_{ABC})\geq E^{(2)}(\rho_{X|YZ})\geq
E^{(2)}(\rho_{XY})$, where XYZ is any permutation of ABC).
Generally, for a $k$-partite entanglement measure $E^{(k)}$, the
unification condition ensures that one can restrict $E^{(k)}$ to any
subsystems and any subpartitions without causing any trouble; the
hierarchy condition mainly requires that, as a kind of physical
resource, the partial entanglement is never greater than the whole
entanglement. Therefore, the unification condition and the hierarchy
condition are natural requirements for $E^{(k)}$ to be a $k$-partite
entanglement measure. But the situation is much more complicated for
$k>3$, particularly, no exact definition for the hierarchy condition
is known.

We remark here that, the hierarchy condition lives also in bipartite
entanglement measure.  In fact,  the inequality $E(\rho_{A|BC})\geq
E(\rho_{AB})$ may be regarded as the hierarchy condition  and should
be satisfied by the bipartite entanglement measure $E$.

As many quantum correlations beyond  entanglement are also physical
resources, naturally,
 when quantifying these multipartite quantum correlations,
 the unification condition and the hierarchy condition
should  be basic requirements. Consider a multipartite quantum
correlation (MQC) and assume that, for any $k\geq 2$, $\mathcal
C^{(k)}$ is a $k$-partite quantum correlation measure for { MQC}.
The meaning of the unification condition is well understood. We say
that $\mathcal C^{(k)}$ satisfies the unification condition if, for
any $2\leq l\leq k$, one has a uniform way to introduce the
$l$-partite quantum correlation measure $\mathcal C^{(l)}$  for any
$l$-partition so that the elements in the sequence $\{\mathcal
C^{(l)}\}_{l=2}^k$ get along well with each other. For the hierarchy
condition, we should consider at least three situations. Roughly
speaking,  $\mathcal C^{(k)}$ satisfies the hierarchy condition
means that, for any multipartite state, the correlation of
subpartition is not greater than the whole correlation;  the
correlation of part is not greater than the correlation of whole;
``kick some parties out of'' each subgroup will not increase the
correlation between subgroups of subsystems.

In the present paper, as $\mathcal M^{(k)}$ is symmetric, we mainly
consider {\it the symmetric multipartite quantum correlations} ({SMQCs}), namely, the multipartite quantum correlations which are
invariant under any permutations of subsystems.

To consider the hierarchy condition for $\mathcal M^{(k)}$ with
$k>2$, one has to check several kinds of inequalities. To make the
question more clear, let us consider the case  of $k=4$. For a $4$-partite
system ABCD, it has three kinds of $2$-subpartitions: $WX|YZ$,
$W|XYZ$, $WXY|Z$; and three kinds of $3$-partitions: $W|X|YZ$,
$W|XY|Z$, $WX|Y|Z$, where $WXYZ$ is any permutation of $ABCD$. So
the hierarchy condition  requires that ${\mathcal M}^{(4)}$ should
satisfy the natural property ``total correlation'' $\geq$ ``partial
correlation'', that is,
$$\begin{array}{rl}{\mathcal M}^{(4)}(\rho_{ABCD})\geq &{\mathcal
M}^{(2)}(\rho_{WX|YZ})\\
\geq& {\mathcal M}^{(2)}(\rho_{WX|Y})\geq
{\mathcal M}^{(2)}(\rho_{X|Y}),\end{array} \eqno(2)
$$
$$\begin{array}{rl}{\mathcal M}^{(4)}(\rho_{ABCD})\geq &{\mathcal
M}^{(2)}(\rho_{W|XYZ})\\
\geq & {\mathcal M}^{(2)}(\rho_{W|XY})\geq
{\mathcal M}^{(2)}(\rho_{W|X}),\end{array}   \eqno(3)
$$
$$\begin{array}{rl}{\mathcal M}^{(4)}(\rho_{ABCD})\geq &{\mathcal
M}^{(3)}(\rho_{W|X|YZ})\\
 \geq &\mathcal M^{(3)}(\rho_{WXY})\geq
\mathcal M^{(2)}(\rho_{W|XY})\end{array}  \eqno(4)
$$
and
$${\mathcal M}^{(4)}(\rho_{ABC|DE})\geq{\mathcal M}^{(4)}(\rho_{ABCD}).  \eqno(5)
$$
Ineqs.(2-5) motivate  in general what kinds of inequalities should be
checked for ${\mathcal M}^{(4)}$ to obey  the hierarchy condition.
Obviously, the truth of Ineqs.(2-4) also implies that both $\mathcal
M^{(3)}$ and $\mathcal M^{(2)}$ meet the hierarchy condition.

Since the unification condition is usually clear from the definition
of a multipartite quantum correlation measure,  we give here only an
exact definition of the hierarchy condition for quantifying
SMQCs.

Let $k\geq 3$. For any $l$-subpartition ${\mathcal P}_l(A_1A_2\ldots
A_k)$ ($2\leq l\leq k$) of $k$-partite system $A_1A_2\ldots A_k$, it
is obvious that there exists  a permutation $\pi$ of $(1,2,\ldots
k)$ and positive integers $i_1,i_2,\ldots, i_l$ with
$i_1+\cdots+i_l=k$ such that
\begin{widetext}
$$\begin{array}{rl} &{\mathcal P}_l(A_1A_2\ldots A_k) \\
= &{\small  A_{\pi(1)}\ldots A_{\pi(i_1)}|A_{\pi(i_1+1)}\ldots
A_{\pi(i_1+i_2)}|A_{\pi(i_1+i_2+1)}\ldots\ldots A_{\pi(i_1+\cdots
+i_{l-1})}|A_{\pi(i_1+\cdots +i_{l-1}+1)}\ldots A_{\pi(k)}}.
\end{array} \eqno(6)$$
\end{widetext}

\begin{definition} For $k\geq 2$, let $\mathcal C^{(k)}$ be a candidate as a $k$-partite quantum
correlation measure of a symmetric multipartite quantum correlation
which satisfies the unification condition on $k$-partite composite
system $A_1A_2\ldots A_k$. Let ${\mathcal P}_l(A_1A_2\ldots A_k) $
 be any $l$-subpartition of $A_1A_2\ldots
A_k$ determined by $2\leq l\leq k$ and the permutation $\pi$ of
$(1,2,\ldots, k)$ as in Eq.(6).

 {\rm (1)} $\mathcal C^{(k)}$ is nonincreasing under subpartition if,
 for any $2\leq l\leq k$ and the permutation $\pi$,
 $${\mathcal C}^{(k)}(\rho_{A_1,A_2,\ldots, A_k})
\geq  {\mathcal C}^{(l)}(\rho_{{\mathcal P}_l(A_1A_2\ldots A_k)})
$$
holds   for any state $\rho_{A_1A_2\ldots A_k}$.

{\rm (2)}  $\mathcal C^{(k)}$ is nonincreasing under taking subgroup
if, for any $2\leq l\leq k$, the permutation $\pi$ and each $h\in\{0,1,
2,\ldots, l\}$ with $i_0=0$,
$$\begin{array}{rl}
& {\mathcal C}^{(k)}(\rho_{A_1,A_2,\ldots, A_k})\\
\geq & {\mathcal C}^{(i_h)}(\rho_{A_{\pi(i_0+i_1+\cdots
+i_{h-1}+1)}\ldots A_{\pi(i_0+i_1+\cdots +i_h)}})
\end{array}
$$
holds   for any state $\rho_{A_1A_2\ldots A_k}$.

{\rm (3)} $\mathcal C^{(l)}$ ($l\geq 2$) is nonincreasing under
kickout, if, for any $k>l$, any permutation $\pi$ of $(1,2,\ldots,
k)$ and for each $h=0,1,2,\ldots, l$, letting $C_h$ be a nonempty
subset of $B_h=\{ A_{\pi((\sum_{j=0}^{h-1} i_j)+1)},
A_{\pi((\sum_{j=0}^{h-1} i_j)+2)},\ldots, A_{\pi(\sum_{j=0}^{h} i_j
)}\}$,
$$\begin{array}{rl}
{\mathcal C}^{(l)}(\rho_{{\mathcal P}_l(A_1A_2\ldots A_k)}) \geq {\mathcal
C}^{(l)}(\rho_{C_1|C_2|\ldots |C_{l-1}|C_l})
\end{array}
$$
holds  for any state $\rho_{A_1A_2\ldots A_k}$.

We say that $\mathcal C^{(k)}$ satisfies the hierarchy condition if
$\mathcal C^{(k)}$ is nonincreasing under subpartition,
nonincreasing under taking subgroup and nonincreasing under kickout.

In the case that $\mathcal C^{(k)}$ is a $k$-partite Gaussian
quantum correlation measure, $\mathcal C^{(k)}$ satisfies the
hierarchy condition if it is nonincreasing under subpartition,
nonincreasing under taking subgroup and nonincreasing under kickout
at least for all $k$-partite Gaussian states.
\end{definition}

The hierarchy condition for bipartite quantum correlation measures
beyond entanglement are less studied. Generally speaking, many known
bipartite Gaussian correlations are not hierarchied. This is often
the case for those quantum correlations with ancilla problem. For
example, considering the Gaussian nonlocality $\mathcal N$
proposed in \cite{WHQ}, with $\rho_{ABC}=\rho_{AB}\otimes \rho_C$,
we have
$$ \mathcal N(\rho_{A|BC})=\mathcal N(\rho_{AB}){\rm
tr}\rho_{C}^{2}<\mathcal N(\rho_{AB})$$ whenever $\rho_C$ is not
pure. This means that  partial correlation may be bigger than the
whole correlation, that is, the hierarchy condition is broken by
$\mathcal N$.  But $\mathcal M(=\mathcal M^{(2)})$ proposed in
\cite{LHQ} satisfies the hierarchy condition:

{\it For any $(m+n+l)$-mode tripartite state $\rho_{ABC}\in{\mathcal
{FS}}(H_A\otimes H_B\otimes H_C)$, we have ${\mathcal
M}(\rho_{A|BC})\geq{\mathcal M}(\rho_{AB})$.}

We claim that ${\mathcal M}^{(4)}$ satisfies
Ineqs.(2-5) and thus meets the hierarchy condition. In fact, this is a
special case of the following general result.

\begin{theorem}
For $k\geq 2$, ${\mathcal M}^{(k)} $ in Definition 1 satisfies the
unification condition and the hierarchy condition. Thus, ${\mathcal
M}^{(k)} $ is a $k$-partite multimode Gaussian quantum correlation
measure.
\end{theorem}

It is clear from the Definition 1 that ${\mathcal M}^{(k)} $
satisfies the unification condition. To show that ${\mathcal
M}^{(k)} $  meets the hierarchy condition, one has to check that, by
Definition 2, $\mathcal M^{(k)}$ is  nonincreasing under
subpartition, nonincreasing under taking subgroup and nonincreasing
under kickout for all $k$-partite ${\mathcal {FS}}$ states. A
proof will be given in  Appendix B.

\section{Monogamy relations for ${\mathcal M}^{(k)}$}

 An
important feature of many bipartite entanglement measures is that
they are monogamous. Here,  we accept a slightly more general concept of
monogamy relation of entanglement without inequalities from \cite{GuoZhang, GG} rather than  the CKW inequality, which says that, if two parties A and
B are maximally entangled, then neither  A nor B can be entangled
with the third party C. In \cite{GuoZhang},
multipartite entanglement measures and multipartite monogamy
relations (mainly tripartite systems) were discussed. For  a
bipartite entanglement measure $E: {\mathcal S}(H_A\otimes H_B)\to
[0,+\infty)$, the monogamy of $E$ implies $E(\rho_{A|BC})\geq
E(\rho_{AB})$,  and $E(\rho_{A|BC})=E(\rho_{AB})$ will force
$E(\rho_{AC})=E(\rho_{BC})=0$ (which is equivalent to the statement
that there exists some $\alpha>0$ such that
$E(\rho_{A|BC})^\alpha\geq E(\rho_{AB})^\alpha+E(\rho_{AC})^\alpha $
\cite{GuoZhang}). Clearly, the monogamy relation of entanglement
accords with the resource allocation theory: if the first part and
the second part share all resources, then the third part can  share
any resource with neither the first part nor the second part.
However, as pointed out in \cite{GuoZhang}, the monogamy relation of
the above form discusses the entanglement allocation among three
parties by a bipartite entanglement measure $E$ and thus is not
complete, because $E(\rho_{A|BC})$ is only a part of the whole
entanglement contained in $\rho_{ABC}$ (or, shared by A,B,C).
Therefore, to discuss the monogamy relation of entanglement, one
needs the help of multipartite entanglement measures.

Different from the bipartite monogamy relation, there are three
kinds of monogamy relations for a tripartite entanglement measure:
$E^{(3)}$ is bipartite like monogamous if
$E^{(3)}(\rho_{A|B|(CD)})=E^{(3)}(\rho_{ABC})$ implies
$E^{(2)}(\rho_{AD})=E^{(2)}(\rho_{BD})=E^{(2)}(\rho_{CD})=0$ (this
monogamy relation is not proposed and discussed in \cite{GuoZhang});
$E^{(3)}$ is completely monogamous if
$E^{(3)}(\rho_{ABC})=E^{(2)}(\rho_{AB})$ implies
$E^{(2)}(\rho_{AC})=E^{(2)}(\rho_{BC})=0$; $E^{(3)}$ is tightly
 monogamous if $E^{(3)}(\rho_{ABC})=E^{(2)}(\rho_{A|BC})$
implies $E^{(2)}(\rho_{BC})=0$. The bipartite like monogamy relation
is more stringent or stronger than the complete monogamy relation
as we always have
$E^{(4)}(\rho_{ABCD})\geq E^{(3)}(\rho_{A|B|CD})$
by the hierarchy condition. So we may call
it the {\it strong monogamy relation}. Thus, the monogamous bipartite entanglement measures are
in fact strongly monogamous.

Note that many multipartite quantum correlations are physical
resources. Naturally, when discussing multipartite quantum correlation
measures, the three kinds of monogamy relations similar to those
mentioned in the previous paragraph for $E^{(3)}$ should be
explored. Let us give a precise definition of the monogamy relations
as follows.

\begin{definition} For $k\geq 2$, let $\mathcal C^{(k)}$ be a $k$-partite quantum
correlation measure of a symmetric multipartite quantum correlation
on a $k$-partite composite system $A_1A_2\ldots A_k$. Let ${\mathcal
P}_l(A_1A_2\ldots A_k)$   be a $l$-subpartition of $A_1A_2\ldots
A_k$ determined by $2\leq l\leq k$ and the permutation $\pi$ of
$(1,2,\ldots, k)$ as in Eq.(6).

{\rm (1) (Tight monogamy relation)} $\mathcal C^{(k)}$ ($k\geq 3$)
is tightly monogamous if, for any $ l$, any $\pi$ and any $k$-partite
state $\rho_{A_1A_2\ldots A_k}$, $$\small {\mathcal
C}^{(k)}(\rho_{A_1A_2\ldots A_k})={\mathcal
C}^{(l)}(\rho_{{\mathcal
P}_l(A_1A_2\ldots A_k)})$$ will imply that, for every $h=1,2,\ldots,l$,
${\mathcal C}^{(i_h)}(\rho_{B_h})=0$ whenever $i_h\geq 2$, where
$B_h=A_{\pi(i_0+i_1+\cdots i_{h-1}+1)}A_{\pi(i_0+i_1+\cdots
i_{h-1}+2)}\ldots A_{\pi(i_0+i_1+\cdots i_{h-1}+i_h)}$ with $i_0=0$.

{\rm (2) (Complete monogamy relation)} $\mathcal C^{(k)}$ ($k\geq
3$) is completely monogamous if, for any $ l$, any $\pi$ with
$i_1\geq2$ and any $k$-partite state $\rho_{A_1A_2\ldots A_k}$,
${\mathcal C}^{(k)}(\rho_{A_1A_2\ldots A_k})={\mathcal
C}^{(i_1)}(\rho_{A_{\pi(1)}A_{\pi(2)}\ldots A_{\pi(i_1)}})$ will
imply that
$${\mathcal C}^{(2)}(\rho_{A_{\pi(1)}\ldots
A_{\pi(i_1)}|A_{\pi(i_1+1)}A_{\pi(i_1+2)}\ldots A_{\pi(k)}})=0 $$
and
$${\mathcal C}^{(k-i_1)}(\rho_{A_{\pi(i_1+1)}A_{\pi(i_1+2)}\ldots A_{\pi(k)}})=0.$$

{\rm (3) (Strong monogamy relation)} $\mathcal C^{(l)}$ ($l\geq 2$)
is strongly monogamous if, for any $k>l$ and for each $h=1,2,\ldots,
l$, letting $C_h$ be a nonempty subset of $B_h=\{
A_{\pi((\sum_{j=0}^{h-1} i_j)+1)}, A_{\pi((\sum_{j=0}^{h-1}
i_j)+2)},\ldots, A_{\pi(\sum_{j=0}^{h} i_j )}\}$,
$$\begin{array}{rl}
{\mathcal C}^{(l)}(\rho_{{\mathcal
P}_l(A_1A_2\ldots A_k)})= {\mathcal
C}^{(l)}(\rho_{C_1|C_2|\ldots |C_{l-1}|C_l})
\end{array}
$$
will imply that ${\mathcal C}^{(r_h)}(\rho_{C_h})=0$ whenever
$r_h\geq 2$ with $r_h$ the number of subsystems contained in
$C_h$, and that ${\mathcal C}^{(2)}(\rho_{A_iA_j})=0$ whenever one of
$A_i$ and $A_j$ is not in $\cup _{h=1}^l C_h$.

If ${\mathcal C}^{(k)}$ is a multipartite multimode Gaussian quantum
correlation measure for CV systems, we require that ${\mathcal
C}^{(k)}$ meets the definition  at least on  Gaussian states.
\end{definition}

Roughly speaking,  {\it the tight monogamy relation} means
that, if the correlation of subpartition attains the total
correlation, then the parties in the same subgroup are not
correlated to each other; {\it the complete monogamy relation} means
that, if the correlation of a subgroup of subsystems attains the
total correlation, then the parties out of the subgroup are not
correlated with any other parties in the system; {\it the strong
monogamy relation}  claims that, if the correlation of subpartition
keeps invariant after ``kicking some parties out of'' each
subgroups, then the remain parties are not correlated with the
parties kicked out of.

The following bipartite like monogamy relation is a  special case of
the strong monogamy relation.

{\rm (4) ({\it Special case of strong monogamy relation})} $\mathcal
C^{(k)}$ ($k\geq 2$) is strongly monogamous if, for any
$(k+1)$-partite state $\rho_{A_1A_2\ldots A_kA_{k+1}}$, ${\mathcal
C}^{(k)}(\rho_{A_1A_2\ldots A_{k-1}|(A_kA_{k+1})})={\mathcal
C}^{(k)}(\rho_{A_1A_2\ldots A_{k-1}A_k})$ will imply that ${\mathcal
C}^{(2)}(\rho_{A_1A_{k+1}})={\mathcal
C}^{(2)}(\rho_{A_2A_{k+1}})=\ldots={\mathcal
C}^{(2)}(\rho_{A_kA_{k+1}})=0$.

 Many bipartite GQCMs
 beyond entanglement are  not monogamous.  For example,   the
Gaussian nonclassicality $\mathcal N$ proposed in \cite{WHQ} is
obviously  not monogamous  since it breaks the hierarchy condition.
Though $\mathcal M=\mathcal M^{(2)}$ obeys the hierarchy condition,
it is not monogamous by the next theorem. Hence it is reasonable
that a good multipartite (Gaussian) quantum correlation measure
should be at least  completely monogamous and tightly monogamous.

\begin{theorem}
The bipartite Gaussian quantum correlation measure $\mathcal M$
 is not (strongly) monogamous.
\end{theorem}

 {\bf Proof.} Let $\rho_{ABC}$ be an $(m+n+l)$-mode
tripartite state with CM
$\Gamma_{ABC}=\left(\begin{array}{ccc} A & X& Z \\ X^T & B & Y \\
Z^T & Y^T & C \end{array}\right)$. Then, by Theorem B3 in Appendix
B, ${\mathcal M}(\rho_{A|BC})= {\mathcal M}(\rho_{AB})$ if and only
if $ XB^{-1}Y=Z$, which may not be zero. However, ${\mathcal
M}(\rho_{AC})={\mathcal M}(\rho_{BC})=0$ if and only if $Z=0$ and
$Y=0$. Thus, ${\mathcal M}(\rho_{A|BC})= {\mathcal M}(\rho_{AB})$
does not imply that ${\mathcal M}(\rho_{AC})={\mathcal
M}(\rho_{BC})=0$.

To make it clearer, we give an   example which reveals that there do
exist tripartite Gaussian state $\rho_{ABC}$ so that ${\mathcal
M}(\rho_{A|BC})={\mathcal M}(\rho_{AB})$ but ${\mathcal
M}(\rho_{AC})\not=0$ and ${\mathcal M}(\rho_{BC})\not=0$. Therefore,
${\mathcal M}$ is not monogamous.

 Let
$$ \tiny\Gamma=\left(\begin{array}{cc|cc|cc}
2 & 0 & 1 & 0&  \frac{1}{3} & 0 \\
0 & 2 & 0 & 1 & 0 & \frac{1}{3} \\
\hline
1 & 0 & 3 & 0& 1 & 0 \\
0 & 1 & 0 & 3 & 0 & 1 \\
\hline
\frac{1}{3} & 0 & 1 & 0 & 2 & 0 \\
0 & \frac{1}{3} & 0 & 1 & 0 & 2
\end{array}\right) \quad{\rm and}\quad
\Delta=\left(\begin{array}{cc|cc|cc}
0& 1 & 0 & 0&  0 & 0 \\
-1 & 0 & 0 & 0 & 0 & 0 \\
\hline
0 & 0 & 0 & 1 & 0 & 0 \\
0 & 0 & -1 & 0 & 0 & 0 \\
\hline
0 & 0 & 0 & 0 & 0 & 1 \\
0 & 0 & 0 & 0 & -1 & 0
\end{array}\right).
$$
Since $\Gamma +i\Delta \geq 0$,  $\Gamma$ is a CM of some
$(1+1+1)$-mode Gaussian state $\rho_{ABC}$.
Note that
$$
\tiny Z=\left(\begin{array}{cc} \frac{1}{3} & 0 \\ 0 & \frac{1}{3}
\end{array}\right)=\left(\begin{array}{cc} 1 & 0 \\ 0 & 1
\end{array}\right)\left(\begin{array}{cc} 3 & 0 \\ 0 & 3
\end{array}\right)^{-1}\left(\begin{array}{cc} 1 & 0 \\ 0 & 1
\end{array}\right)=XB^{-1}Y.
$$
However, it is easily calculated that ${\mathcal M}(\rho_{A|BC})={\mathcal
M}(\rho_{AB})\approx 0.3056$, ${\mathcal
M}(\rho_{AC})\approx 0.0548\not=0$ and ${\mathcal M}(\rho_{BC}) \approx 0.3056\not=0$.
So $\mathcal M$
breaks the monogamy relation at $\rho_{ABC}$. \hfill$\Box$

The fact that $\mathcal M$ is not monogamous is not surprising
because,  by the hierarchy condition, ${\mathcal
M}(\rho_{A|BC})=0.3056$ is just a part of the total quantum
correlation ${\mathcal M}^{(3)}(\rho_{ABC})=0.5144$ shared by three
parties A, B, C. So ${\mathcal M}(\rho_{A|BC})={\mathcal
M}(\rho_{AB})$ cannot always force that both ${\mathcal
M}(\rho_{AC})$ and ${\mathcal M}(\rho_{BC})$ are zero.

However, for some special
tripartite Gaussian states $\sigma_{ABC}$, ${\mathcal
M}(\sigma_{A|BC})={\mathcal M}(\sigma_{AB})$ do imply that
${\mathcal M}(\sigma_{AC})={\mathcal M}(\sigma_{BC})=0$.

{\bf Example 1.} For any tripartite fully symmetric Gaussian state
$\sigma_{ABC}$, ${\mathcal M}(\sigma_{A|BC})={\mathcal
M}(\sigma_{AB})$ implies that ${\mathcal M}(\sigma_{AC})={\mathcal
M}(\sigma_{BC})=0$.  In fact we have  ${\mathcal
M}(\sigma_{A|BC})={\mathcal M}(\sigma_{AB})$ if and only if
$\sigma_{ABC}=\sigma_A\otimes \sigma_B\otimes \sigma_C$.

Recall that $\sigma_{ABC}$ is fully symmetric if it is an
$(n+n+n)$-mode Gaussian state with  CM having the form
$$
\Gamma_{\sigma_{ABC}}=\left(\begin{array}{ccc} A & X & X \\ X^T & A
& X
\\ X^T& X^T & A
\end{array}\right),
$$
where $A,X\in M_{2n}(\mathbb R)$. We have to show that ${\mathcal
M}(\sigma_{A|BC})={\mathcal M}(\sigma_{AB})$ if and only if $X=0$.
This will be done in Appendix C.

To illustrate the meaning of three kinds of monogamy relations
in Definition 3 for multipartite multimode GQCM $\mathcal C^{(k)}$, we consider the case $k=4$. For
a $4$-partite composite system ABCD, let $WXYZ$ be any permutation
of $ABCD$. The three kinds of monogamy relations for ${\mathcal
C}^{(4)}$ can be stated as follows.

(a) (Tight monogamy relation) $\mathcal C^{(4)}$ is
tightly monogamous if

(a$_1$) ${\mathcal C}^{(4)}(\rho_{ABCD})= {\mathcal
C}^{(2)}(\rho_{WX|YZ})$
  implies that ${\mathcal C}^{(2)}(\rho_{WX})=
{\mathcal C}^{(2)}(\rho_{YZ})=0, $

(a$_2$) ${\mathcal C}^{(4)}(\rho_{ABCD})={\mathcal
C}^{(2)}(\rho_{W|XYZ})$ implies that ${\mathcal C}^{(3)}(\rho_{XYZ})=0
$ \\
and

(a$_3$) ${\mathcal C}^{(4)}(\rho_{ABCD})= {\mathcal
C}^{(3)}(\rho_{W|X|YZ})$ implies that $ {\mathcal C}^{(2)}(\rho_{YZ})=0.
$

(b) (Complete monogamy relation) $\mathcal C^{(4)}$ is completely
monogamous if

(b$_1$) $ {\mathcal C}^{(4)}(\rho_{ABCD})= {\mathcal
C}^{(3)}(\rho_{WXY})$ implies that
${\mathcal C}^{(2)}(\rho_{WZ})=
{\mathcal C}^{(2)}(\rho_{XZ})= {\mathcal C}^{(2)}(\rho_{YZ})=0
$ \\ and

(b$_2$)  $ {\mathcal C}^{(4)}(\rho_{ABCD})= {\mathcal
C}^{(2)}(\rho_{WX})$
implies that ${\mathcal C}^{(2)}(\rho_{WX|YZ})=
{\mathcal C}^{(2)}(\rho_{YZ})=0.$

(c) (A special case of the strong monogamy relation) $\mathcal
C^{(4)}$ is strongly monogamous if
$ {\mathcal M}^{(4)}(\rho_{ABC|(DE)})= {\mathcal
M}^{(4)}(\rho_{ABCD})$ implies that ${\mathcal M}^{(2)}(\rho_{AE})=
{\mathcal M}^{(2)}(\rho_{BE})
 = {\mathcal
M}^{(2)}(\rho_{CE})={\mathcal M}^{(2)}(\rho_{DE})=0.
$

We claim that ${\mathcal M}^{(4)}$  is completely monogamous and tightly
monogamous, but is not strongly monogamous. In fact, this is
a special case of the following general result, which will be
proved in Appendix C.

\begin{theorem} Let ${\mathcal M}^{(k)}: {\mathcal
{FS}}(H_{A_1}\otimes H_{A_2}\otimes\cdots\otimes H_{A_k}) \to
[0,+\infty)$ be the $k$-partite $(n_1+n_2+\cdots+n_k)$-mode Gaussian
quantum correlation in Definition 1.

{\rm (1)} For $k\geq 3$, ${\mathcal M}^{(k)}$
 is tightly monogamous.

 {\rm (2)} For $k\geq 3$, ${\mathcal M}^{(k)}$
 is completely monogamous.

 {\rm (3)} For $k\geq 2$, ${\mathcal M}^{(k)}$
 is not strongly monogamous.
\end{theorem}

{\bf Example 2.} Consider the $(k+1)$-partite $(1+\cdots+1)$-mode
case. Let $\Gamma=(A_{ij})$ be a real symmetric matrix, where
$A_{jj}=\left(\begin{array}{cc} 2 &0
\\ 0 & 2 \end{array}\right)$ for $j=1,2,\ldots, k-1,k+1$, $A_{kk}=\left(\begin{array}{cc} 3 &0
\\ 0 & 3 \end{array}\right)$, $A_{k-1,k}=A_{k,k+1}=\left(\begin{array}{cc} 1 &0
\\ 0 & 1 \end{array}\right)$, $A_{k-1,k+1}=\left(\begin{array}{cc} \frac{1}{3} &0
\\ 0 & \frac{1}{3} \end{array}\right)$, and otherwise, $A_{ij}=0$ for
  $i<j$. It is easily checked that $\Gamma=\Gamma_\rho$ is a CM for some Gaussian state
  $\rho=\rho_{A_1A_2\ldots A_kA_{k+1}}$ since $\Gamma+i \oplus_{j=1}^{k+1}\Delta_j\geq 0$
  with $\Delta_j=\left(\begin{array}{cc}
  0&1
\\ -1 & 0 \end{array}\right)$. For this
state $\rho_{A_1A_2\ldots A_{k-1}A_kA_{k+1}}$, we have ${\mathcal
M}^{(k)}(\rho_{A_1A_2\ldots A_{k-1}|A_kA_{k+1}})={\mathcal
M}^{(k)}(\rho_{A_1A_2\ldots A_{k-1}A_k})=0.3056$, but $\mathcal
M^{(2)}(\rho_{A_{k-1}A_{k+1}})\approx 0.0548\not=0$. Hence $\mathcal
M^{(k)}$ is not strongly monogamous.

\section{Conclusion}

For a $k$-partite $(n_1+n_2+\cdots+n_k)$-mode Gaussian state
$\rho=\rho_{A_1A_2\ldots A_k}$, we say that $\rho$ is not quantum
correlated if it is a $k$-partite product state, that is
$\rho=\rho_{A_1}\otimes \rho_{A_2}\otimes\cdots\otimes\rho_{A_k}$.
In this paper, we propose a computable multipartite multimode
Gaussian quantum correlation measure ${\mathcal M}^{(k)}$ for any
$k$-partite multimode continuous-variable (CV) systems. The value of ${\mathcal
M}^{(k)}$ only depends on the covariance matrices of  CV states, is
invariant under any permutation of subsystems, has no ancilla
problem, is nonincreasing under $k$-partite local Gaussian channels
(particularly, invariant under $k$-partite local Gaussian unitary
operations), and vanishes on $k$-partite product states. For a
$k$-partite Gaussian state $\rho$, ${\mathcal M}^{(k)}(\rho)=0$ if
and only if $\rho$ is a $k$-partite product state. Moreover, as a
multipartite Gaussian quantum correlation measure,
${\mathcal M}^{(k)}$ satisfies the unification condition and the hierarchy
condition that a multipartite quantum correlation measure should obey
(which means that, ${\mathcal M}^{(k)}$ is consistent with
${\mathcal M}^{(l)}$ for any $2\leq l\leq k$,
 the correlation of  subpartition is not greater than the whole
correlation,  the correlation of part is not greater than the
correlation of whole, and  the correlation after
 kicking some parties out of  subgroups is not greater than the correlation between the subgroups).
 Finally, the monogamy relations for
multipartite quantum
 correlation measures are discussed. Generally speaking, there are three kinds of
 monogamy relations for a multipartite correlation measure: the strong monogamy relation (the correlation between
subgroups after ``kicking some parties out of'' each subgroups keeps
invariant will imply that the remain parties are not correlated with
the parties kicked out of), the complete
 monogamy relation (the correlation of a
subgroup attains the total correlation will imply that the parties
out of the subgroup are not correlated with any other parties of the
system) and the tight monogamy condition (the correlation
between subgroups attains the total correlation will imply that the
parties in the same subgroup are not correlated to each other).
Though  ${\mathcal M}^{(k)}$ is not strongly monogamous, ${\mathcal
M}^{(k)}$ is completely monogamous and tightly monogamous. Thus
${\mathcal M}^{(k)}$ is a nice multipartite multimode Gaussian quantum
correlation measure. As $\mathcal M^{(k)}$ is easily calculated,  it
is more convenient  to be applied in other scenarios of quantum
information.

By now, we think that $\mathcal M^{(k)}$ is the only known multipartite
multimode Gaussian quantum correlation measure beyond
entanglement. It is interesting to find
other multipartite multimode Gaussian quantum correlation measures.

{\bf Acknowledgement.}  The authors wish to give their thanks to the
referees. They read the original manuscript carefully and gave many
helpful comments to improve the paper.
This work is supported by the National
Natural Science Foundation of China (12071336, 12171290) and Fund Program for the Scientific
Activities of Selected Returned Overseas Professionals in Shanxi
Province (20200011).

\begin{widetext}
\vspace{1mm}
\renewcommand{\thesection}{APPENDIX}

\section{}

\subsection{Proofs of basic properties of ${\mathcal M}^{(k)}$}

This appendix section is devoted to proving  the properties 4)-6) of
${\mathcal M}^{(k)}$ in Section III.

{\bf Property 4)}.  ${\mathcal M}^{(k)}$ is invariant under
$k$-partite local Gaussian unitary operation.

{\bf Proof.} For an $n$-mode CV system determined by ${R}=(\hat
R_1,\hat R_2,\cdots,\hat
R_{2n})=(\hat{Q}_1,\hat{P}_1,\cdots,\hat{Q}_n,\hat{P}_n)$,  it is known that a unitary operator $U$ is Gaussian if and only if
there is a vector $\mathbf m$   in ${\mathbb R}^{2n}$ and a   matrix
$\mathbf S \in {\rm Sp}(2n,\mathbb R)$ such that $U^{\dag}RU = \mathbf SR +\mathbf m$ (\cite{Wang,Christian}), where
${\rm Sp}(2n,\mathbb R)$ is
the symplectic group of all
$2n\times 2n$ real matrices $\bf S$ that satisfy
$\mathbf S\in{\rm Sp}(2n,\mathbb R)\Leftrightarrow \mathbf S\Delta \mathbf S^{\rm
T}=\Delta$.
Thus, every Gaussian unitary operator $U$ is determined
by some affine symplectic map $(\mathbf S, \mathbf m)$ acting on the
phase space, and can be parameterized as $U=U_{\mathbf S, \mathbf m}$. It follows that, if $U_{\mathbf S,\mathbf m}$ is a Gaussian unitary
operator, then, for  any $n$-mode
 state $\rho$   with CM $\Gamma_\rho$ and mean $\mathbf d_\rho$,
the  state $\sigma=U_{\mathbf S,\mathbf m}\rho U_{\mathbf
S,\mathbf m}^{\dag}$ has the CM $\Gamma_{\sigma} = \mathbf S\Gamma
\mathbf S^{\rm T}$ and the mean $\mathbf d_{\sigma} =\mathbf m + \mathbf
S\mathbf d$. Particularly, if $\rho$ is also Gaussian, then the
characteristic function of the Gaussian state $\sigma$ is of the form
$\exp(-\frac{1}{4}z^{\rm T}\Gamma_{\sigma} z+i\mathbf
d_{\sigma}^{\rm T}z)$.

Now, assume that
$\rho=\rho_{A_1,A_2,\ldots,A_k}\in{\mathcal{FS}}(H_{A_1}\otimes
H_{A_2}\otimes\cdots\otimes H_{A_k})$ is an
$(n_1+n_2+\cdots+n_k)$-mode $k$-partite state, and $U_{\mathbf S,
\mathbf m}=U_1\otimes U_2\otimes\cdots \otimes U_k$ is a Gaussian
unitary operator with $U_j=U_{\mathbf S_j, \mathbf m_j}$ being
Gaussian unitary on $H_{A_j}$. Clearly, $\mathbf S=\oplus_{j=1}^k
\mathbf S_j$ and $\mathbf m=\oplus_{j=1}^k \mathbf m_j$. Let
$\sigma=\sigma_{A_1,A_2,\ldots,A_k}=U_{\mathbf S, \mathbf
m}\rho_{A_1,A_2,\ldots,A_k}U_{\mathbf S, \mathbf m}^\dag$. Then $\Gamma_\sigma=\mathbf S\Gamma_\rho \mathbf S^T$
and $\Gamma_{\sigma_{A_j}}=\mathbf S_j\Gamma_{\rho_{A_j}}\mathbf
S_j^T$. As $\det(\mathbf S)=\Pi_{j=1}^k\det(\mathbf S_j)$, it
follows from Definition 1 that
$$\begin{array}{rl} \mathcal
M^{(k)}(\sigma_{A_1,A_2,\ldots,A_k})=&
1-\frac{\det(\Gamma_\sigma)}{\Pi_{j=1}^k\det(\Gamma_{\sigma_{A_j}})}=1-\frac{\det(\mathbf
S\Gamma_\rho\mathbf S^T)}{\Pi_{j=1}^k\det(\mathbf
S_j\Gamma_{\rho_{A_j}}\mathbf S_j^T)} \\ = & 1-\frac{\det(\mathbf
S)\det(\Gamma_\rho)\det(\mathbf S^T)}{\Pi_{j=1}^k\det(\mathbf
S_j)\det(\Gamma_{\rho_{A_j}})\det(\mathbf
S_j^T)}=1-\frac{\det(\Gamma_\rho)}{\Pi_{j=1}^k\det(\Gamma_{\rho_{A_j}})}
\\ =& {\mathcal M}^{(k)}(\rho_{A_1,A_2,\ldots,A_k}),
\end{array}$$
as desired. \hfill$\Box$

{\bf Property 5)}. For any $(n_1+n_2+\cdots +n_k)$-mode $k$-partite
state $\rho_{A_1,A_2,\ldots, A_k}\in{\mathcal {FS}}(H_{A_1}\otimes
H_{A_2}\otimes\cdots \otimes H_{A_k})$ with CM
$\Gamma=(A_{ij})_{k\times k}$ as in Eq.(1), $\mathcal
M^{(k)}(\rho_{A_1,A_2,\ldots, A_k})=0$ if and only if $A_{ij}=0$
whenever $i\not=j$. Particularly, if $\rho_{A_1,A_2,\ldots, A_k}$ is
a Gaussian state, then $\mathcal M^{(k)}(\rho_{A_1,A_2,\ldots,
A_k})=0$ if and only if $\rho_{A_1,A_2,\ldots, A_k}$ is a
$k$-partite product Gaussian state, that is, $\rho_{A_1,A_2,\ldots,
A_k}=\rho_{A_1}\otimes \rho_{A_2}\otimes\cdots \otimes \rho_{A_k}$.

{\bf Proof.} This is an immediate consequence of  Lemma A1
below.\hfill$\Box$

{\bf Lemma A1.} {\it Assume that
$$\Gamma_k=\left(\begin{array}{cccc} A_{11} & A_{12} &\cdots & A_{1k}\\
A_{21} & A_{22} &\cdots & A_{2k} \\ \vdots & \vdots & \ddots &\vdots
\\ A_{k1} & A_{k2} & \cdots & A_{kk}  \end{array}\right)
$$
is a   positive definite block matrix over the complex field
$\mathbb C$. Then $\det (\Gamma_k) =\Pi_{j=1}^k\det(A_{jj})$ if and
only if $A_{ij}=0$ whenever $i\not= j$.}

To prove Lemma A1, we need the following lemma  proved in \cite{LHQ}
which is also useful in the other part of the present paper:

{\bf Lemma A2.}  {\it For $S,T\in M_{n}({\mathbb C})$ with $S\geq T>
0$, $\det (S)=\det(T)$ if and only if $S=T$.}

{\bf Proof of Lemma A1.} The ``if" part is obvious. We prove the
``only if" part by induction on $k$. For the case $k=2$, denote
$\Gamma_2 =\left(\begin{array}{cc}
A & C \\
C^{T} & B
\end{array}
\right)$. It is well known that $\det(\Gamma_2)=\det(A)\det(B-C^{\rm
T}A^{-1}C)=\det(B)\det(A-CB^{-1}C^T)>0$. If $\det(\Gamma_2)=\det (A)
\det (B)$, then $\det(B-C^{\rm T}A^{-1}C)=\det B$. Let
$D=B-C^{\rm T}A^{-1}C$. As $\Gamma_2>0$, we have  $0< D\leq B$.
Thus, by Lemma A2 we must have $B=D$ as $\det(D)=\det(B)$, which
entails that $C=0$.

Now, assume that the assertion is true
for $k-1\geq 2$. Denote by $\Gamma_{k-1}$ the  principal submatrix of $\Gamma_k$, that is, $$\Gamma_{k-1}=\left(\begin{array}{cccc} A_{11} & A_{12} &\cdots & A_{1,k-1}\\
A_{21} & A_{22} &\cdots & A_{2,k-1} \\ \vdots & \vdots & \ddots
&\vdots
\\ A_{k-1,1} & A_{k-1,2} & \cdots & A_{k-1,k-1}  \end{array}\right),
$$
which is positive definite, too. Note that the condition
$\det(\Gamma_k)=\Pi_{j=1}^k\det(A_{jj})$ implies
$$\begin{array}{rl}
& \Pi_{j=1}^k\det(A_{jj})= \det(\Gamma_k)
\\ = & \det(A_{kk})\det(\Gamma_{k-1}-\left(\begin{array}{c} A_{1k}  \\
A_{2k}  \\ \vdots \\ A_{k-1,k}
\end{array}\right) A_{kk}^{-1} \left(\begin{array}{cccc} A_{1k}^\dag &
A_{2k}^\dag & \cdots & A_{k-1,k}^\dag
\end{array}\right)) \\
\leq & \det(A_{kk})\det(\Gamma_{k-1})\leq \det(A_{kk}) \Pi_{j=1}
^{k-1}\det(A_{jj})=\Pi_{j=1}^k\det(A_{jj}).
\end{array}\eqno(A1)
$$
It follows that $\Pi_{j=1}^{k-1}\det(A_{jj})=  \det(\Gamma_{k-1})$.
By the inductive assumption, $A_{ij}=0$ whenever
$i\not=j$ and $i,j\in\{1,2,\ldots, k-1\}$. The remain is to check
that $A_{jk}=0$ for all $j=1,2,\ldots, k-1$. By Eq.(A1) again, one
gets
$$
\det(\Gamma_{k-1}-\left(\begin{array}{c} A_{1k}  \\  A_{2k}  \\
\vdots \\ A_{k-1,k}
\end{array}\right) A_{kk}^{-1} \left(\begin{array}{cccc} A_{1k}^\dag &
A_{2k}^\dag & \cdots & A_{k-1,k}^\dag
\end{array}\right))=\det(\Gamma_{k-1}),
$$
which forces
$$
\left(\begin{array}{c} A_{1k}  \\  A_{2k}  \\
\vdots \\ A_{k-1,k}
\end{array}\right) A_{kk}^{-1}=0,
$$
and so $A_{jk}=0$ for all $j=1,2,\ldots, k-1$. Hence the ``only if"
part is also true, completing the proof. \hfill$\Box$

{\bf Property 6)}. (Nonincreasing under local Gaussian channels)
For any Gaussian state $\rho_{A_1,A_2,\ldots, A_k}\in{\mathcal
{FS}}(H_{A_1}\otimes H_{A_2}\otimes\cdots\otimes H_{A_k})$ and any
local Gaussian channel $\Phi_1\otimes
\Phi_2\otimes\cdots\otimes\Phi_k$, we have
$$\mathcal M^{(k)}((\Phi_1\otimes \Phi_2\otimes\cdots\otimes\Phi_k)\rho_{A_1,A_2,\ldots, A_k})\leq \mathcal M^{(k)}(\rho_{A_1,A_2,\ldots, A_k}).$$
Particularly, ${\mathcal M}^{(k)}$ is locally Gaussian unitary
invariant.

To prove Property 6), we need a lemma on matrices from \cite{LHQ}:

{\bf Lemma A3.} {\it Let $B,K,M\in M_n(\mathbb C)$ with $B$ and $M$
positive semidefinite.  If both $B$ and $KBK^\dag+M$ are invertible,
then
$K^\dag(KBK^\dag +M)^{-1} K\leq B^{-1}$.
The equality holds if and only if $M=0$ and $K$ is invertible. }

{\bf Proof of Property 6).} As Gaussian state $\rho$ is characerized by
its CM $\Gamma$ and mean $\mathbf d$, we can parameterize
it as $\rho=\rho(\Gamma,\mathbf d)$. Recall that, if $\Phi$ is a
Gaussian channel of $n$-mode Gaussian systems, then, for any $n$-mode Gaussian state
$\rho=\rho(\Gamma,\mathbf d)$,
$\Phi(\rho(\Gamma,\mathbf
d))=\rho(\Gamma^{\prime},\mathbf d^{\prime})$ with
$$\mathbf d^{\prime}=K\mathbf d+\overline{\mathbf d}\ \ {\rm and}\ \ \Gamma^{\prime}=K\Gamma K^{T}+M \eqno(A2)$$
for some
real matrices $M, K\in M_{2n}(\mathbb R)$ satisfying $M=M^T\geq 0$
and det$ M\geq (\det (K)-1)^2$, and some vector  $\overline{\mathbf
d}\in {\mathbb R}^{2n}$. So we can parameterize the
Gaussian channel $\Phi$ as $\Phi=\Phi(K, M, \overline{\mathbf d})$.

Let $\rho=\rho_{A_1,A_2,\ldots, A_k}\in{\mathcal
{FS}}(H_{A_1}\otimes H_{A_2}\otimes\cdots\otimes H_{A_k})$ be a
Gaussian state whose CM is presented as in Eq.(1) and $\Phi_1\otimes
\Phi_2\otimes\cdots\otimes\Phi_k$ be a local Gaussian channel with
$\Phi_j=\Phi_j(K_j, M_j, \overline{\mathbf d}_j)$. We first show
that, for any $j\in\{1,2,\ldots, k\}$,
$$\mathcal M^{(k)}((I_1\otimes \cdots I_{j-1}\otimes \Phi_j\otimes I_{j+1}\otimes\cdots\otimes I_k)\rho_{A_1,A_2,\ldots, A_k})\leq \mathcal M^{(k)}(\rho_{A_1,A_2,\ldots, A_k}),\eqno(A3)$$
where $\Phi_j$ is a Gaussian channel performed on subsystem A$_j$.
Since ${\mathcal M}^{(k)}$ is invariant under permutation of
subsystems, we may assume that $j=k$.  Denote by
$$\rho'=\rho'_{A_1,A_2,\ldots,
A_k}=(I_1\otimes I_2\otimes\cdots\otimes
I_{k-1}\otimes\Phi_k)\rho_{A_1,A_2,\ldots, A_k}.$$ Then the
CM $\Gamma_{\rho'}$ of  $\rho'_{A_1,A_2,\ldots, A_k}$
has the form
$$\Gamma_{\rho'}=\left(\begin{array}{ccccc} A_{11} & A_{12} &\cdots & A_{1,k-1} & A_{1k}K_k^T\\
A_{21} & A_{22} &\cdots & A_{2,k-1} & A_{2k}K_k^T \\ \vdots & \vdots
& \ddots &\vdots & \vdots \\  A_{k-1,1} & A_{k-1,2} & \cdots &
A_{k-1,k-1} & A_{k-1,k} \\
 K_kA_{k1} & K_kA_{k2} & \cdots & K_kA_{k,k-1} & K_kA_{kk}K_k^T +M_k \end{array}\right),
$$ and thus, by Lemma A3, one gets
$$ \begin{array}{rl}
& \mathcal M^{(k)}(\rho'_{A_1,A_2,\ldots,
A_k})=1-\frac{\det(\Gamma_{\rho'})}{\det(K_kA_{kk}K_k^T
+M_k)\Pi_{j=1}^{k-1}\det(A_{jj})} \\
= & 1-\frac{\det(\Gamma_{k-1}-\left(\begin{array}{c} A_{1k} \\
A_{2k} \\ \vdots \\ A_{k-1,k}
\end{array}\right) K_k^T(K_kA_{kk}K_k^T +M_k)^{-1}K_k\left(\begin{array}{cccc} A_{1k}^T & A_{2k}^T &\cdots & A_{k-1,k}^T
\end{array}\right))}{\Pi_{j=1}^{k-1}\det(A_{jj})}\\
\leq & 1-\frac{\det(\Gamma_{k-1}-\left(\begin{array}{c} A_{1k} \\
A_{2k} \\ \vdots \\ A_{k-1,k}
\end{array}\right)A_{kk}^{-1}\left(\begin{array}{cccc} A_{1k}^T & A_{2k}^T &\cdots & A_{k-1,k}^T
\end{array}\right))}{\Pi_{j=1}^{k-1}\det(A_{jj})} =\mathcal M^{(k)}(\rho_{A_1,A_2,\ldots,
A_k}).
 \end{array}$$
 Therefore, we have proved that the inequality in Eq.(A3) is true.

Then, applying the inequality in Eq.(A3), we have
$$\begin{array}{rl}
& \mathcal M^{(k)}((\Phi_1\otimes
\Phi_2\otimes\cdots\otimes\Phi_k)\rho_{A_1,A_2,\ldots, A_k})\\
= & \mathcal M^{(k)}((\Pi_{j=1}^k (I_1\otimes \cdots I_{j-1}\otimes \Phi_j\otimes I_{j+1}\otimes\cdots\otimes I_k))\rho_{A_1,A_2,\ldots, A_k}) \\
\leq & \mathcal M^{(k)}((\Pi_{j=2}^k (I_1\otimes \cdots
I_{j-1}\otimes \Phi_j\otimes I_{j+1}\otimes\cdots\otimes
I_k))\rho_{A_1,A_2,\ldots, A_k}) \\
\leq & \ldots\ldots \\
 \leq & \mathcal M^{(k)}((I_1\otimes \cdots
I_{k-1}\otimes \Phi_k)\rho_{A_1,A_2,\ldots, A_k}) \\
 \leq & \mathcal
M^{(k)}(\rho_{A_1,A_2,\ldots, A_k}).
\end{array}$$
Hence, ${\mathcal M}^{(k)}$ is nonincreasing under $k$-partite local
Gaussian channels.

Particularly, if the  $k$-partite local Gaussian channel
$\Phi=\Phi_1\otimes \Phi_2\otimes \cdots\otimes \Phi_k$ is
invertible and $\Phi^{-1}$ is still a Gaussian channel (it is the
case when $\Phi$ is a $k$-partite locally Gaussian unitary operation),
then
$$ {\mathcal M}^{(k)}(\rho_{A_1A_2\ldots A_k})= {\mathcal M}^{(k)}(\Phi^{-1}\Phi(\rho_{A_1A_2\ldots
A_k}))\leq  {\mathcal M}^{(k)}(\Phi(\rho_{A_1A_2\ldots A_k}))\leq
{\mathcal M}^{(k)}(\rho_{A_1A_2\ldots A_k}),
$$
and consequently, ${\mathcal M}^{(k)}(\Phi(\rho_{A_1A_2\ldots
A_k}))= {\mathcal M}^{(k)}(\rho_{A_1A_2\ldots A_k})$. This reveals
that $ {\mathcal M}^{(k)} $ is invariant under $k$-partite Gaussian
unitary operations. \hfill$\Box$

\subsection{Proof of Theorem 1, the hierarchy condition for ${\mathcal M}^{(k)}$}

In the appendix section, we show that $\mathcal M^{(k)}$ satisfies
the hierarchy condition, and thus complete the proof of Theorem 1 in
Section IV.

We begin with considering bipartite case. The following lemmas are
needed.

{\bf Lemma B1.}
{\it Let $\left(\begin{array}{ccc} I & D& F \\ D^\dag &I  & E \\
F^\dag & E^\dag & I \end{array}\right)$ be a positive definite block
matrix over the complex field $\mathbb C$. Then
$\max\{\|D\|,\|E\|,\|F\|\} <1$,
$$ \left(\begin{array}{cc} D & F \end{array}\right)\left(\begin{array}{cc}    I & E \\
  E^\dag & I \end{array}\right)^{-1}\left(\begin{array}{c} D^\dag \\
  F^\dag \end{array}\right)\geq DD^\dag,\ \ \left(\begin{array}{cc} D & F \end{array}\right)\left(\begin{array}{cc}    I & E \\
  E^\dag & I \end{array}\right)^{-1}\left(\begin{array}{c} D^\dag \\
  F^\dag \end{array}\right)\geq FF^\dag,
$$
$$ \left(\begin{array}{cc} F^\dag & E^\dag \end{array}\right)\left(\begin{array}{cc}    I & D \\
  D^\dag & I \end{array}\right)^{-1}\left(\begin{array}{c} F \\
 E \end{array}\right)\geq F^\dag F
\ \ {\rm and}\ \ \left(\begin{array}{cc} F^\dag & E^\dag \end{array}\right)\left(\begin{array}{cc}    I & D \\
  D^\dag & I \end{array}\right)^{-1}\left(\begin{array}{c} F \\
 E \end{array}\right)\geq E^\dag E.
$$
Furthermore, the equality holds for any one of the above four
inequalities if and only if the equality holds for all of the above
four inequalities, and in turn, if and only if $F=DE$. }

{\bf Proof.} By the assumption, $\left(\begin{array}{ccc} I & D& F \\ D^\dag &I  & E \\
F^\dag & E^\dag & I \end{array}\right)\geq 0$ and is invertible. So
we must have  $\max\{\|D\|,\|E\|,\|F\|\} <1$. We only give a proof
of the first inequality in detail, the others are checked similarly
by noting that
$$
E(I-E^\dag E)^{-1}=(I-EE^\dag )^{-1}E \quad {\rm and}\quad (I-E^\dag
E)^{-1} E^\dag=E^\dag (I-EE^\dag)^{-1}.
$$
It is easily checked that
$$\left(\begin{array}{cc}    I & E \\
  E^\dag & I \end{array}\right)^{-1}=\left(\begin{array}{cc}  (I-EE^\dag)^{-1} & -E(I-E^\dag E)^{-1} \\
-(I-E^\dag E)^{-1}E^\dag & (I-E^\dag E)^{-1} \end{array}\right).
$$
Then,
$$\begin{array}{rl} & \left(\begin{array}{cc} D & F\end{array}\right)\left(\begin{array}{cc}    I & E \\
  E^{\dag} & I \end{array}\right)^{-1}\left(\begin{array}{c} D^{\dag} \\
  F^{\dag} \end{array}\right) \\
  =&
   D(I-EE^{\dag})^{-1}D^{\dag}-F(I-E^{\dag}E)^{-1}E^{\dag}D^{\dag}-DE(I-E^{\dag}E)^{-1}F^{\dag}+F(I-E^{\dag}E)^{-1}F^{\dag}\\
  =& DD^{\dag}+DE(I-E^{\dag}E)^{-1}E^{\dag}D^{\dag}-F(I-E^{\dag}E)^{-1}E^{\dag}D^{\dag}\\
  & -DE(I-E^{\dag}E)^{-1}F^{\dag}+F(I-E^{\dag}E)^{-1}F^{\dag} \\
  = & DD^{\dag}+ \left(\begin{array}{cc} D & F\end{array}\right)\left(\begin{array}{cc}  E(I-E^{\dag}E)^{-1}E^{\dag} & -E(I-E^{\dag}E)^{-1} \\
-(I-E^{\dag}E)^{-1}E^{\dag} & (I-E^{\dag}E)^{-1} \end{array}\right)\left(\begin{array}{c} D^{\dag} \\
  F^{\dag} \end{array}\right).
  \end{array}
$$
Obviously,
$$\left(\begin{array}{cc}  E(I-E^{\dag}E)^{-1}E^{\dag} & -E(I-E^{\dag}E)^{-1} \\
-(I-E^{\dag}E)^{-1}E^{\dag} & (I-E^{\dag}E)^{-1}
\end{array}\right)\geq 0
$$
since
$(-E(I-E^{\dag}E)^{-1})[(I-E^{\dag}E)^{-1}]^{-1}(-(I-E^{\dag}E)^{-1}E^{\dag}
)=E(I-E^{\dag}E)^{-1}E^{\dag}$. Hence we have
$$\begin{array}{rl} & \left(\begin{array}{cc} D & F \end{array}\right)\left(\begin{array}{cc}    I & E \\
  E^{\dag} & I \end{array}\right)^{-1}\left(\begin{array}{c} D^{\dag} \\
  F^{\dag} \end{array}\right) \\
  = &  DD^{\dag}+\left(\begin{array}{cc} D & F \end{array}\right)\left(\begin{array}{cc}  E(I-E^{\dag}E)^{-1}E^{\dag} & -E(I-E^{\dag}E)^{-1} \\
-(I-E^{\dag}E)^{-1}E^{\dag} & (I-E^{\dag}E)^{-1} \end{array}\right)\left(\begin{array}{c} D^{\dag} \\
  F^{\dag} \end{array}\right)\geq DD^{\dag},
  \end{array} \eqno(B1) $$
as desired.

It is clear from Ineq.(B1)   that the equality holds if and only if
$$
\left(\begin{array}{cc} D & F \end{array}\right)\left(\begin{array}{cc}  E(I-E^{\dag}E)^{-1}E^{\dag} & -E(I-E^{\dag}E)^{-1} \\
-(I-E^{\dag}E)^{-1}E^{\dag} & (I-E^{\dag}E)^{-1} \end{array}\right)\left(\begin{array}{c} D^{\dag} \\
  F^{\dag} \end{array}\right)=0.
$$
As, for operators $A,C$ with  $A\geq 0$, $CAC^\dag=0 \Leftrightarrow
CA=0$, we see that the above equation holds if and only if
$$\begin{array}{l} DE(I-E^\dag E)^{-1}E^\dag -F(I-E^\dag)^{-1}E^\dag=0,\\
  -DE(I-E^\dag E)^{-1} +F(I-E^\dag E)^{-1} =0;
  \end{array}
$$
and in turn, if and only if $F=DE$.

It is similar to show that the equality for any one of the other
three inequalities holds if and only if the same condition $F=DE$ is
satisfied, completing the proof. \hfill$\Box$

The following lemma is a generalization of Lemma B1, which is also
useful.

{\bf Lemma B2.}
{\it Let $\left(\begin{array}{ccc} A & X& Z \\ X^\dag &B  & Y \\
Z^\dag & Y^\dag & C \end{array}\right)$ be a positive definite block
matrix over the complex field $\mathbb C$. Then
$$ \left(\begin{array}{cc} X & Z \end{array}\right)\left(\begin{array}{cc}    B & Y \\
  Y^\dag & C \end{array}\right)^{-1}\left(\begin{array}{c} X^\dag \\
  Z^\dag \end{array}\right)\geq XB^{-1}X^\dag,
\ \ \left(\begin{array}{cc} X & Z \end{array}\right)\left(\begin{array}{cc}   B & Y \\
  Y^\dag & C \end{array}\right)^{-1}\left(\begin{array}{c} X^\dag \\
  Z^\dag \end{array}\right)\geq ZC^{-1}Z^\dag,
$$
$$ \left(\begin{array}{cc} Z^\dag & Y^\dag \end{array}\right)\left(\begin{array}{cc}    A & X \\
  X^\dag & B \end{array}\right)^{-1}\left(\begin{array}{c} Z \\
 Y \end{array}\right)\geq Z^\dag A^{-1}Z
\ \ {\rm and}\ \  \left(\begin{array}{cc} Z^\dag & Y^\dag \end{array}\right)\left(\begin{array}{cc}    A & X \\
  X^\dag & B \end{array}\right)^{-1}\left(\begin{array}{c} Z \\
 Y \end{array}\right)\geq Y^\dag B^{-1}Y.
$$
Furthermore, the equality holds for any one of the above four
inequalities if and only if the equality holds for all of the above
four inequalities, and in turn, if and only if $Z=XB^{-1}Y$. }

{\bf Proof.} Clearly, $$ \Gamma=\left(\begin{array}{ccc} A & X& Z \\ X^\dag &B  & Y \\
Z^\dag & Y^\dag & C \end{array}\right)=\left(\begin{array}{ccc} A^{\frac{1}{2}} & 0& 0 \\  0 & B^{\frac{1}{2}} & 0 \\
0 & 0 & C^{\frac{1}{2}} \end{array}\right)\left(\begin{array}{ccc} I & D& F \\ D^\dag &I  & E \\
F^\dag & E^\dag & I \end{array}\right)\left(\begin{array}{ccc} A^{\frac{1}{2}} & 0& 0 \\  0 & B^{\frac{1}{2}} & 0 \\
0 & 0 & C^{\frac{1}{2}} \end{array}\right),$$ where
$D=A^{-\frac{1}{2}}XB^{-\frac{1}{2}}$,
$E=B^{-\frac{1}{2}}YC^{-\frac{1}{2}}$ and
$F=A^{-\frac{1}{2}}ZC^{-\frac{1}{2}}$. As $\Gamma$ is positive and
invertible, we have $\{\|D\|, \|E\|,\|F\|\}\subset [0, 1)$. Then
Lemma 4 is applicable. Let us give a proof of the second inequality
in detail. By Lemma 4 we have
$$ \left(\begin{array}{cc} D & F \end{array}\right)\left(\begin{array}{cc}    I & E \\
  E^\dag & I \end{array}\right)^{-1}\left(\begin{array}{c} D^\dag \\
  F^\dag \end{array}\right)\geq FF^\dag.
$$
Substituting $D=A^{-\frac{1}{2}}XB^{-\frac{1}{2}}$,
$E=B^{-\frac{1}{2}}YC^{-\frac{1}{2}}$ and
$F=A^{-\frac{1}{2}}ZC^{-\frac{1}{2}}$ into the above inequality
leads to
$$\begin{array}{rl}
 A^{-\frac{1}{2}}\left(\begin{array}{cc} X & Z \end{array}\right)\left(\begin{array}{cc}   B & Y \\
  Y^\dag & C \end{array}\right)^{-1}\left(\begin{array}{c} X^\dag \\
  Z^\dag \end{array}\right)A^{-\frac{1}{2}}=\left(\begin{array}{cc} D & F \end{array}\right)\left(\begin{array}{cc}    I & E \\
  E^\dag & I \end{array}\right)^{-1}\left(\begin{array}{c} D^\dag \\
  F^\dag \end{array}\right)
  \geq  FF^\dag =A^{-\frac{1}{2}}ZC^{-1}Z^\dag A^{-\frac{1}{2}},
  \end{array}
$$
which entails that
$$ \left(\begin{array}{cc} X & Z \end{array}\right)\left(\begin{array}{cc}   B & Y \\
  Y^\dag & C \end{array}\right)^{-1}\left(\begin{array}{c} X^\dag \\
  Z^\dag \end{array}\right)\geq ZC^{-1}Z^\dag.
$$
By Lemma B1, the equality holds if and only if $F=DE$, which holds
if and only if $Z=XB^{-1}Y$, completing the proof.\hfill$\Box$

The next result reveals that the bipartite Gaussian quantum
correlation measure $\mathcal M=\mathcal M^{(2)}$ satisfies the
hierarchy condition.

 {\bf Theorem B3.} {\it
For any $(m+n+l)$-mode tripartite state $\rho_{ABC}\in{\mathcal
{FS}}(H_A\otimes H_B\otimes H_C)$, we have ${\mathcal
M}(\rho_{A|BC})\geq{\mathcal M}(\rho_{AB})$. Furthermore, ${\mathcal
M}(\rho_{A|BC})={\mathcal M}(\rho_{AB})$ if and only if
$Z=XB^{-1}Y$, where  $\Gamma_{ABC}=\left(\begin{array}{ccc} A & X& Z \\ X^T & B & Y \\
Z^T & Y^T & C \end{array}\right)$ is the covariance matrix of
$\rho_{ABC}$. }

 {\bf Proof.} Let $\rho_{ABC}$ be an $(m+n+l)$-mode
tripartite state with CM
$\Gamma_{ABC}=\left(\begin{array}{ccc} A & X& Z \\ X^T & B & Y \\
Z^T & Y^T & C \end{array}\right)$. Then the CM of $\rho_A$ is $A$
and
the CM of $\rho_{BC}$ is $\Gamma_{BC}=\left(\begin{array}{cc}    B & Y \\
  Y^T & C \end{array}\right)$.  Clearly,
$$\Gamma_{ABC}=\left(\begin{array}{ccc} A^{\frac{1}{2}} & 0& 0 \\  0 & B^{\frac{1}{2}} & 0 \\
0 & 0 & C^{\frac{1}{2}} \end{array}\right)\left(\begin{array}{ccc} I & D& F \\ D^T &I  & E \\
F^T & E^T & I \end{array}\right)\left(\begin{array}{ccc} A^{\frac{1}{2}} & 0& 0 \\  0 & B^{\frac{1}{2}} & 0 \\
0 & 0 & C^{\frac{1}{2}} \end{array}\right),$$ where
$D=A^{-\frac{1}{2}}XB^{-\frac{1}{2}}$,
$E=B^{-\frac{1}{2}}YC^{-\frac{1}{2}}$ and
$F=A^{-\frac{1}{2}}ZC^{-\frac{1}{2}}$. As $\Gamma_{ABC}$ is
invertible, we have $\{\|D\|, \|E\|,\|F\|\}\subset [0, 1)$. It
follows that
$$\begin{array}{rl} {\mathcal
M}(\rho_{A|BC})=& 1-\frac{\det(\Gamma_{ABC})}{\det(A)\det(\Gamma_{BC})}=1-\frac{\det(\left(\begin{array}{ccc} I & D& F \\ D^T &I  & E \\
F^T & E^T & I \end{array}\right))}{\det(\left(\begin{array}{cc}    I & E \\
  E^T & I \end{array}\right))}= 1-\det(I-(D\ F)\left(\begin{array}{cc}    I & E \\
  E^T & I \end{array}\right)^{-1}\left(\begin{array}{c} D^T \\
  F^T \end{array}\right))
  \end{array}
$$
and
$${\mathcal
M}(\rho_{AB})=1-\frac{\det(\Gamma_{AB})}{\det(A)\det(B)}=1-\det(I-D^TD)=1-\det(I-DD^T).
$$
Thus,  by Lemma B1, we have
$$  (D\ F)\left(\begin{array}{cc}    I & E \\
  E^T & I \end{array}\right)^{-1}\left(\begin{array}{c} D^T \\
  F^T \end{array}\right)\geq  DD^T.
$$
Hence
$$ \det(I-(D\ F)\left(\begin{array}{cc}    I & E \\
  E^T & I \end{array}\right)^{-1}\left(\begin{array}{c} D^T \\
  F^T \end{array}\right))\leq \det(I-DD^T),
$$
and consequently ${\mathcal M}(\rho_{A|BC})\geq {\mathcal
M}(\rho_{AB})$. 

Now, by   Lemma A2, it is easily seen that  ${\mathcal
M}(\rho_{A|BC})= {\mathcal M}(\rho_{AB})$ if and only if
$$\left(\begin{array}{cc} D & F \end{array}\right)\left(\begin{array}{cc}    I & E \\
  E^T & I \end{array}\right)^{-1}\left(\begin{array}{c} D^T \\
  F^T \end{array}\right)=DD^T.
$$
Therefore, by Lemma B1, we conclude that  ${\mathcal
M}(\rho_{A|BC})= {\mathcal M}(\rho_{AB})$ if and only if $F=DE$,
which is equivalent to  say that $ XB^{-1}Y=Z$.
 This completes the proof.
\hfill$\Box$

Now let us consider the general case.

{\bf Theorem B4.}  {\it Let $\rho_{A_1,A_2,\ldots, A_k}\in{\mathcal
{FS}}(H_{A_1}\otimes H_{A_2}\otimes\cdots\otimes H_{A_k})$ with the
CM as in Eq.(1). For any $l$-subpartition  ($2\leq l< k$)   of
$k$-partite system ($k\geq 3$) $A_1A_2\ldots A_k$ as in Eq.(6),  the
following statements are true.

 {\rm (i) (Nonincreasing under subpartition)} $$\begin{array}{rl}
 & {\mathcal M}^{(k)}(\rho_{A_1,A_2,\ldots, A_k})
\geq  {\mathcal M}^{(l)}(\rho_{A_{\pi(1)}\ldots
A_{\pi(i_1)}|A_{\pi(i_1+1)}\ldots A_{\pi(i_1+i_2)}|\ldots\ldots
|A_{\pi(i_1+\cdots +i_{l-1}+1)}\ldots A_{\pi(i_1+\cdots +i_l)}}).
\end{array}
$$

{\rm (ii)  (Nonincreasing under taking subgroup)} With $i_0=0$, for
each $h\in\{1, 2,\ldots, l\}$, we have
$$\begin{array}{rl}
& {\mathcal M}^{(k)}(\rho_{A_1,A_2,\ldots, A_k})
\geq {\mathcal M}^{(i_h)}(\rho_{A_{\pi(i_0+i_1+\cdots
+i_{h-1}+1)}\ldots A_{\pi(i_0+i_1+\cdots +i_h)}}).
\end{array}
$$
}

{\bf Proof.} (i) Let $\rho_{A_1,A_2,\ldots, A_k}\in{\mathcal
{FS}}(H_{A_1}\otimes H_{A_2}\otimes\cdots\otimes H_{A_k})$ with the
CM as in Eq.(1) and ${\mathcal P}_l(A_1A_2\ldots
A_k)$ be an $l$-partition of $A_1A_2\ldots A_k$ as in Eq.(6). Denote
by $B_h=A_{\pi(i_0+i_1+\cdots i_{h-1}+1})\cdots
A_{\pi(i_0+i_1+\cdots i_{h-1}+i_h)}$ and $\Gamma_{B_h}$ the CM of
$\rho_{B_h}$, $h=1,2,\ldots, k$. Then, by Definition 1 and Lemma A1,
we have, with $i_0=0$,
$$
\begin{array}{rl}
& {\mathcal M}^{(l)}(\rho_{A_{\pi(1)}\ldots
A_{\pi(i_1)}|A_{\pi(i_1+1)}\ldots A_{\pi(i_1+i_2)}|\ldots
|A_{\pi(i_1+\cdots +i_{l-1}+1)}\ldots A_{\pi(k)}})\\
=& 1-\frac{\det(\Gamma_{\rho_{A_1A_2\ldots A_k}})}{\Pi_{h=1}^l
\det(\Gamma_{B_h})}
 \leq  1-\frac{\det(\Gamma_{\rho_{A_1A_2\ldots A_k}})}{\Pi_{h=1}^l (\Pi_{j=i_0+i_1+\cdots+i_{h-1}+1}^{i_1+\cdots+i_{h-1}+i_h}\det(A_{\pi(j),\pi(j)}))}
 = {\mathcal M}^{(k)}(\rho_{A_1A_2\ldots A_k}).
\end{array} \eqno(B2)
$$
This completes the proof of (i).

(ii) For any $h\in\{1,2,\ldots,l\}$, we also denote by
$B_h=\{\pi(i_0+i_1+\cdots +i_{h-1}+1),\ldots, \pi(i_0+i_1+\cdots
+i_{h-1}+i_h) \}$ and $\Gamma_{B_h^c}=(A_{ij})$ with $i,j\in B_h^c=
\{1,2,\ldots, k\}\setminus B_h$, which is the CM of $\rho_{B_h^c}$.
Then
$$\Gamma_\rho\cong\Gamma_{\rho_{A_{\pi(1)}\ldots A_{\pi(i_1)}|A_{\pi(i_1+1)}\ldots
A_{\pi(i_1+i_2)}|\ldots |A_{\pi(i_1+\cdots +i_{l-1}+1)}\ldots
A_{\pi(i_1+\cdots +i_l)}}}\cong \Gamma_{\rho_{B_hB_h^c}}=\left(
\begin{array}{cc} \Gamma_{B_h} & C_h \\ C_h^T & \Gamma_{B_h^c}
\end{array}\right).$$ Thus, for any $h=1,2,\ldots, l$, we have
$$
\begin{array}{rl}
&{\mathcal M}^{(i_h)}(\rho_{A_{\pi(i_0+i_1+\cdots +i_{h-1}+1)}\ldots
A_{\pi(i_0+i_1+\cdots +i_h)}})= 1- \frac{\det(\Gamma_{B_h})}{\Pi_{j=i_0+i_1+\cdots+i_{h-1}+1}^{i_0+i_1+\cdots+i_{h-1}+i_h}\det(A_{\pi(j),\pi(j)})}\\
= &
1-\frac{\det(\Gamma_{B_h})\det(\Gamma_{B_h^c}-C_h^T\Gamma_{B_h}^{-1}C_h)}{\det(\Gamma_{B_h^c}-C_h^T\Gamma_{B_h}^{-1}C_h)\Pi_{j=i_0+i_1+\cdots+i_{h-1}+1}^{i_0+i_1+\cdots+i_{h-1}+i_h}\det(A_{\pi(j),\pi(j)})}
\\ \leq & 1-\frac{\det(\Gamma_{\rho_{A_1A_2\ldots
A_k}})}{\det(\Gamma_{B_h^c})\Pi_{j=i_0+i_1+\cdots+i_{h-1}+1}^{i_0+i_1+\cdots+i_{h-1}+i_h}\det(A_{\pi(j),\pi(j)})}\leq
1-\frac{\det(\Gamma_{\rho_{A_1A_2\ldots
A_k}})}{\Pi_{j=1}^{k}\det(A_{jj})}={\mathcal M}^{(k)}(\rho_{A_1A_2\ldots A_k}).
\end{array} \eqno(B3)
$$
This completes the proof of statement (ii). \hfill$\Box$

To prove that $\mathcal M^{(k)}$ satisfies the
hierarchy condition, we still need   a multipartite version of
Theorem B3.

{\bf Theorem B5.} {\rm (Nonincreasing under kickout)} {\it For any
$l$-subpartition ($2\leq l< k$) of $k$-partite system ($k\geq 3$)
$A_1A_2\ldots A_k$ as in Eq.(6) with $i_l\geq 2$ and any
$\rho_{A_1A_2\ldots A_k}\in{\mathcal {FS}}(H_{A_1}\otimes
H_{A_2}\otimes\cdots\otimes H_{A_k})$, we have
$$\begin{array}{rl}
 & {\mathcal M}^{(l)}(\rho_{A_{\pi(1)}\ldots
A_{\pi(i_1)}|A_{\pi(i_1+1)}\ldots A_{\pi(i_1+i_2)}|\ldots\ldots
|A_{\pi(i_1+\cdots +i_{l-1}+1)}\ldots A_{\pi(k-1)}A_{\pi(k)}})\\
\geq & {\mathcal M}^{(l)}(\rho_{A_{\pi(1)}\ldots
A_{\pi(i_1)}|A_{\pi(i_1+1)}\ldots A_{\pi(i_1+i_2)}|\ldots\ldots
|A_{\pi(i_1+\cdots +i_{l-1}+1)}\ldots A_{\pi(k-1)}}).
\end{array}
$$}

{\bf Corollary B6.}  {\it For $k\geq 2$ and $\rho=\rho_{A_1A_2\ldots
A_kA_{k+1}}\in{\mathcal {FS}}(H_{A_1}\otimes H_{A_2}\otimes
\cdots\otimes H_{A_k}\otimes H_{A_{k+1}})$ with CM
$\Gamma_\rho=(A_{ij})_{(k+1)\times (k+1)}$, we always have
$
{\mathcal M}^{(k)}(\rho_{A_1A_2\ldots A_{k-1}|A_kA_{k+1}})\geq
{\mathcal M}^{(k)}(\rho_{A_1A_2\ldots A_k}).
$
}

Note that $\mathcal M^{(k)}$ is symmetric about subsystems. So
the following corollary is true, which is a generalization of
Theorem B5 and reveals exactly the meaning of ``nonincreasing under
kickout''.

{\bf Corollary B7.} {\rm (Nonincreasing under kickout)} {\it Assume
$k\geq 3$ and consider any $l$-partition ($2\leq l< k$) of
$k$-partite system
 $A_1A_2\ldots A_k$ as in Eq.(5). For each
$h=1,2,\ldots, l$, let $C_h$ be a nonempty subset of $B_h=\{
A_{\pi((\sum_{j=0}^{h-1} i_j)+1)}, A_{\pi((\sum_{j=0}^{h-1}
i_j)+2)},\ldots, A_{\pi(\sum_{j=0}^{h} i_j )}\}$, where $i_0=0$.
Then, for any $\rho_{A_1A_2\ldots A_k}\in{\mathcal
{FS}}(H_{A_1}\otimes H_{A_2}\otimes\cdots\otimes H_{A_k})$, we have
$$\begin{array}{rl}
{\mathcal M}^{(l)}(\rho_{A_{\pi(1)}\ldots
A_{\pi(i_1)}|A_{\pi(i_1+1)}\ldots A_{\pi(i_1+i_2)}|\ldots\ldots
|A_{\pi(i_1+\cdots +i_{l-1}+1)}\ldots A_{\pi(k)}}) \geq {\mathcal
M}^{(l)}(\rho_{C_1|C_2|\ldots |C_{l-1}|C_l}).
\end{array}
$$
}

{\bf Proof of Theorem B5.} By the invariance of $\mathcal M^{(k)}$
under permutations, with no loss of generality, we may assume that
the $l$-subpartition of $A_1A_2\ldots A_{k-1}A_k$ is
$$A_1\ldots
A_{i_1}|A_{i_1+1}\ldots A_{i_1+i_2}|A_{i_1+i_2+1}\ldots\ldots
A_{i_1+\cdots+i_{l-1}}|A_{i_1+\cdots+i_{l-1}+1}\ldots
A_{k-1,k-1}A_{kk}. \eqno(B4)$$ As $i_l\geq 2$, we see that
$i_1+i_2+\cdots+i_{l-1}<k-1$.

For any $k$-partite state $\rho=\rho_{A_1A_2\ldots A_k}\in{\mathcal
{FS}}(H_{A_1}\otimes H_{A_2}\otimes\cdots\otimes H_{A_k})$ with CM
as represented as in Eq.(1), write $B_h$ the $h$th party of the
$l$-subpartition in Eq.(B4), that is, $B_h=A_{i_0+i_1+\cdots
+i_{h-1}+1}\ldots A_{i_1+\cdots i_h}$, and $\rho_{B_h}$ the
corresponding reduced state of $\rho_{A_1A_2\ldots A_k}$,
$h=1,2,\ldots, l$. Let $\rho_{k^c}$ be the   reduced state
$\rho_{k^c}=\rho_{A_1A_2\ldots A_{k-1,k-1}}$ of $\rho_{A_1A_2\ldots
A_k}$. Then
$$
{\mathcal M}^{(l)}(\rho_{A_1\ldots A_{i_1}|A_{i_1+1}\ldots
A_{i_1+i_2}|\ldots \ldots|A_{i_1+\cdots+i_{l-1}+1}\ldots
A_{k-1,k-1}A_{kk}})=1-\frac{\det(\Gamma_{\rho})}{\Pi_{h=1}^l
\det(\Gamma_{\rho_{B_h}})}
$$
and
$$\begin{array}{rl}
& {\mathcal M}^{(l)}(\rho_{A_1\ldots A_{i_1}|A_{i_1+1}\ldots
A_{i_1+i_2}|\ldots \ldots|A_{i_1+\cdots+i_{l-1}+1}\ldots
A_{k-1,k-1}})=
1-\frac{\det(\Gamma_{\rho_{k^c}})}{\det(\Gamma_{\rho_{A_{i_1+\cdots
+i_{l-1}+1}\ldots A_{k-1£¬k-1}}})\Pi_{h=1}^{l-1}
\det(\Gamma_{\rho_{B_h}})}.
\end{array}$$
So the theorem is true if and only if
$$
\frac{\det(\Gamma_{\rho})}{  \det(\Gamma_{\rho_{B_l}})}\leq
\frac{\det(\Gamma_{\rho_{k^c}})}{\det(\Gamma_{\rho_{A_{i_1+\cdots
+i_{l-1}+1}\ldots A_{k-1£¬k-1}}})}. \eqno(B5)
$$
Decompose $\Gamma_\rho$ into
$$
\Gamma_\rho=(A_{ij})=\left(\begin{array}{cccc} A_{11}^{\frac{1}{2}}
& 0 & \cdots &
0 \\ 0 & A_{22}^{\frac{1}{2}} & \cdots & 0 \\
\vdots & \vdots & \ddots & \vdots \\
0 &0 & \cdots & A_{kk}^{\frac{1}{2}} \end{array}\right)
\left(\begin{array}{cccc} I & E_{12} & \cdots &
E_{1k} \\ E_{12}^T & I & \cdots & E_{2k} \\
\vdots & \vdots & \ddots & \vdots \\
E_{1k}^T & E_{2k}^T & \cdots & I
\end{array}\right)\left(\begin{array}{cccc} A_{11}^{\frac{1}{2}} & 0
& \cdots &
0 \\ 0 & A_{22}^{\frac{1}{2}} & \cdots & 0 \\
\vdots & \vdots & \ddots & \vdots \\
0 &0 & \cdots & A_{kk}^{\frac{1}{2}} \end{array}\right),
$$
where $E_{ij}=A_{ii}^{-\frac{1}{2}}A_{ij}A_{jj}^{-\frac{1}{2}}$
($i\not=j$), and denote
$$
\Lambda_\rho=\left(\begin{array}{cccc} I & E_{12} & \cdots &
E_{1k} \\ E_{12}^T & I & \cdots & E_{2k} \\
\vdots & \vdots & \ddots & \vdots \\
E_{1k}^T & E_{2k}^T & \cdots & I
\end{array}\right).
$$
Then the inequality in Eq.(B5) holds if and only if
$$
\frac{\det(\Lambda_{\rho})}{  \det(\Lambda_{\rho_{B_l}})}\leq
\frac{\det(\Lambda_{\rho_{k^c}})}{\det(\Lambda_{\rho_{A_{i_1+\cdots
+i_{l-1}+1}\ldots A_{k-1,k-1}}})}. \eqno(B6)
$$
Rewrite
$$
\Lambda_\rho=\left(\begin{array}{cccc} I & E_{12} & \cdots &
E_{1k} \\ E_{12}^T & I & \cdots & E_{2k} \\
\vdots & \vdots & \ddots & \vdots \\
E_{1k}^T & E_{2k}^T & \cdots & I
\end{array}\right)=\left(\begin{array}{ccc} \Lambda_{\rho_{A_1\ldots A_{i_1+\cdots+i_{l-1}}}}& X & Z \\
X^T & \Lambda_{\rho_{A_{i_1+\cdots +i_{l-1}+1}\ldots A_{k-1}}} &  Y \\
Z^T & Y^T &   I
\end{array}\right),
$$
where
$$
X=\left(\begin{array}{cccc} E_{1,i_1+\cdots +i_{l-1}+1}& E_{1,i_1+\cdots +i_{l-1}+2} & \cdots & E_{1,k-1} \\
E_{2,i_1+\cdots +i_{l-1}+1}& E_{2,i_1+\cdots +i_{l-1}+2} & \cdots & E_{2,k-1} \\
\vdots & \vdots & \ddots & \vdots \\
E_{i_1+\cdots +i_{l-1},i_1+\cdots +i_{l-1}+1}& E_{i_1+\cdots
+i_{l-1},i_1+\cdots +i_{l-1}+2} & \cdots & E_{i_1+\cdots
+i_{l-1},k-1}
\end{array}\right), $$
$$ Z=\left(\begin{array}{c} E_{1k} \\ E_{2k} \\ \vdots \\
E_{i_1+\cdots +i_{l-1},k}
\end{array}\right) \quad{\rm and}\quad Y=\left(\begin{array}{c} E_{i_1+\cdots +i_{l-1}+1,k} \\ E_{i_1+\cdots +i_{l-1}+2,k} \\ \vdots \\ E_{k-1,k}
\end{array}\right).
$$
Then $\Lambda_{\rho_{k^c}}=\left(\begin{array}{cc}\Lambda_{\rho_{A_1\ldots A_{i_1+\cdots+i_{l-1}}}}& X   \\
X^T & \Lambda_{\rho_{A_{i_1+\cdots +i_{l-1}+1}\ldots A_{k-1}}}
\end{array}\right)$,
$$
\det(\Lambda_{\rho})=\det(\Lambda_{\rho_{k^c}})\det(I-\left(\begin{array}{cc}
Z^T& Y^T
\end{array}\right)
\Lambda_{\rho_{k^c}}^{-1}\left(\begin{array}{c} Z \\ Y
\end{array}\right))
$$
and
$$
\det(\Lambda_{\rho_{B_l}})=\det(\Lambda_{\rho_{A_{i_1+\cdots
+i_{l-1}+1}\ldots A_{k-1,k-1}}})\det(I-Y^T
\Lambda_{\rho_{A_{i_1+\cdots +i_{l-1}+1}\ldots A_{k-1,k-1}}}^{-1}
Y).
$$
So the inequality in Eq.(B6) holds if and only if
$$
\det(I-\left(\begin{array}{cc} Z^T& Y^T
\end{array}\right)
\Lambda_{\rho_{k^c}}^{-1}\left(\begin{array}{c} Z \\ Y
\end{array}\right))\leq \det(I-Y^T
\Lambda_{\rho_{A_{i_1+\cdots +i_{l-1}+1}\ldots A_{k-1,k-1}}}^{-1}
Y). \eqno(B7)
$$
Now, by the fourth inequality in Lemma B2, we have
$$
\left(\begin{array}{cc} Z^T& Y^T
\end{array}\right)
\Lambda_{\rho_{k^c}}^{-1}\left(\begin{array}{c} Z \\ Y
\end{array}\right) \geq Y^T
\Lambda_{\rho_{A_{i_1+\cdots +i_{l-1}+1}\ldots A_{k-1,k-1}}}^{-1} Y,
\eqno(B8)
$$
which implies that the inequality in Eq.(B7) is true. Consequently,
the inequality in Eq.(B6) is true, and hence the inequality in
Eq.(B5) holds, which completes the proof. \hfill$\Box$

Now, it is clear from Theorem B4 and Corollary B7 that
$\mathcal M^{(k)}$ satisfies the conditions (1)-(3) in Definition 2 and thus meets the hierarchy condition. This completes the proof of Theorem 1.

\subsection{Proof  of Theorem 3}

We first complete the proof of Example 1, that is, to show that, for
any tripartite fully symmetric Gaussian state $\sigma_{ABC}$,
${\mathcal M}(\sigma_{A|BC})={\mathcal M}(\sigma_{AB})$ if and only
if $\sigma_{ABC}=\sigma_A\otimes \sigma_B\otimes \sigma_C$.

{\bf Proof of Example 1.} Recall that $\sigma_{ABC}$ is fully
symmetric if it is an $(n+n+n)$-mode Gaussian state with  CM having
the form
$$
\Gamma_{\sigma_{ABC}}=\left(\begin{array}{ccc} A & X & X \\ X^T & A
& X
\\ X^T& X^T & A
\end{array}\right),
$$
where $A,X\in M_{2n}(\mathbb R)$. We have to show that ${\mathcal
M}(\sigma_{A|BC})={\mathcal M}(\sigma_{AB})$ if and only if $X=0$.

The ``if'' part is obvious. For the ``only if'' part, assume
${\mathcal M}(\sigma_{A|BC})={\mathcal M}(\sigma_{AB})$. By Theorem
B3,  $X=XA^{-1}X$.

Assume that $X\not= 0$. If $X$ is invertible, then $X=A$, which is
impossible as $ \Gamma_{\sigma_{ABC}}$ is invertible. So $\ker
X\not=\{0\}$. Under the space decomposition ${\mathbb R}^{2n}=\ker
X\oplus (\ker X)^\perp$, $A$ and $X$ can be represented as
$$ A=\left(\begin{array}{cc} A_{11} & A_{12} \\ A_{12}^T & A_{22}
\end{array}\right) \quad{\rm and}\quad X=\left(\begin{array}{cc} 0 & X_{12} \\ 0 & X_{22}
\end{array}\right).
$$
Notice that
$$
A^{-1}=\left(\begin{array}{cc} (A_{11}-A_{12}A_{22}^{-1}A_{12}^T)^{-1} & -A_{11}^{-1} A_{12}(A_{22}-A_{12}^TA_{11}^{-1}A_{12})^{-1} \\
-(A_{22}-A_{12}^TA_{11}^{-1}A_{12})^{-1}A_{12}^TA_{11}^{-1} &
(A_{22}-A_{12}^TA_{11}^{-1}A_{12})^{-1}
\end{array}\right).
$$
Then $X=XA^{-1}X$ gives
$$
X_{22}=0 \quad{\rm and}\quad
X_{12}=-X_{12}(A_{22}-A_{12}^TA_{11}^{-1}A_{12})^{-1}A_{12}^TA_{11}^{-1}X_{12}
$$
with $\ker X_{12}=\{0\}$. It follows that $A_{ij}, X_{ij}\in
M_n(\mathbb R)$ and $A_{12}, X_{12}$ are invertible. Therefore, a
tripartite fully symmetric Gaussian state $\sigma_{ABC}$ satisfies
${\mathcal M}(\sigma_{A|BC})={\mathcal M}(\sigma_{AB})$ if and only
if its CM has the form
$$
\Gamma_{\sigma_{ABC}}=\left(\begin{array}{cc|cc|cc} A_{11} &A_{12}& 0 & X_{12} & 0 & X_{12} \\
A_{12}^T & A_{22} &0 & 0 & 0 &0\\
\hline
0& 0 & A_{11} & A_{12} & 0 & X_{12} \\
X_{12}^T & 0 & A_{12}^T & A_{22} & 0 & 0 \\
\hline 0 & 0 & 0 & 0 & A_{11} & A_{12} \\
X_{12}^T & 0 & X_{12}^T & 0 & A_{12}^T & A_{22}
\end{array}\right),
$$
where $A_{ij}, X_{12}\in M_n(\mathbb R)$ are invertible and
$$X_{12}=-
A_{11}(A_{12}^{-1})^T(A_{22}-A_{12}^TA_{11}^{-1}A_{12}).$$ As $
\Gamma_{\sigma_{ABC}}>0$, we have
$$ \left(\begin{array}{cc} A_{11} & A_{12} \\ A_{12}^T & A_{22}
\end{array}\right) \geq \left(\begin{array}{cc} 0 & X_{12} \\ 0 & 0
\end{array}\right)\left(\begin{array}{cc} A_{11} & A_{12} \\ A_{12}^T & A_{22}
\end{array}\right)^{-1}\left(\begin{array}{cc} 0 & 0 \\ X_{12}^T & 0
\end{array}\right)=\left(\begin{array}{cc} D & 0 \\ 0 & 0
\end{array}\right),
$$
where
$$\begin{array}{rl}D=&X_{12}(A_{22}-A_{12}^TA_{11}^{-1}A_{12})^{-1}X_{12}^T=A_{11}(A_{12}^{-1})^T(A_{22}-A_{12}^TA_{11}^{-1}A_{12})A_{12}^{-1}A_{11}\\
=& A_{11}(A_{12}^{-1})^TA_{22}A_{12}^{-1}A_{11}-A_{11}.\end{array}$$
Thus we have
$$
2A_{11}-A_{11}(A_{12}^{-1})^TA_{22}A_{12}^{-1}A_{11}\geq
A_{12}A_{22}^{-1}A_{12}^T.
$$
Note that, $\left(\begin{array}{cc} A_{11} & A_{12} \\ A_{12}^T &
A_{22}
\end{array}\right)>0
$ implies that there is a  contractive matrix $E$ with $\|E\|<1$
such that $A_{12}=A_{11}^{\frac{1}{2}}EA_{22}^{\frac{1}{2}}$. $E$ is
invertible as $A_{12}$ is. So the above inequality becomes to
$$
2A_{11}\geq
A_{11}^{\frac{1}{2}}EE^TA_{11}^{\frac{1}{2}}+A_{11}^{\frac{1}{2}}(EE^T)^{-1}A_{11}^{\frac{1}{2}},
$$
and consequently,
$$
2I\geq EE^T+(EE^T)^{-1}.
$$
This is impossible because it leads to a contraction $0\geq
(I-EE^T)^2>0$. Therefore,   we must have $X=0$ and
$\sigma_{ABC}=\sigma_A\otimes\sigma_B\otimes\sigma_C$.

 As an illustration,   let us
consider the $(1+1+1)$-mode case. If $\Gamma$ is a CM of an
$(1+1+1)$-mode symmetric Gaussian state $\sigma_{ABC}$ satisfying
${\mathcal M}(\sigma_{A|BC})={\mathcal M}(\sigma_{AB})$ and
$\sigma_{ABC}$ not a product state,  then, by what discussed
above, $\Gamma$ has the form
$$
\Gamma=\left(\begin{array}{cc|cc|cc} a &c & 0 & -\frac{ab-c^2}{c} & 0 & -\frac{ab-c^2}{c} \\
c & b &0 & 0 & 0 &0\\
\hline
0& 0 & a & c & 0 & -\frac{ab-c^2}{c} \\
-\frac{ab-c^2}{c} & 0 & c & b & 0 & 0 \\
\hline 0 & 0 & 0 & 0 & a & c \\
-\frac{ab-c^2}{c} & 0 & -\frac{ab-c^2}{c} & 0 & c & b
\end{array}\right),
$$
where $a,b,c\in\mathbb R$ with $a>0, b>0$ and  $ab> c^2>0$.  Since
$$
\Gamma +i\Delta=\left(\begin{array}{cc|cc|cc} a &c+i & 0 & -\frac{ab-c^2}{c} & 0 & -\frac{ab-c^2}{c} \\
c-i & b &0 & 0 & 0 &0\\
\hline
0& 0 & a & c+i & 0 & -\frac{ab-c^2}{c} \\
-\frac{ab-c^2}{c} & 0 & c-i & b & 0 & 0 \\
\hline 0 & 0 & 0 & 0 & a & c+i \\
-\frac{ab-c^2}{c} & 0 & -\frac{ab-c^2}{c} & 0 & c-i & b
\end{array}\right)\geq 0,
$$
we have
$$
ab-c^2-1\geq 0
$$
and
$$
(ab-c^2-1)^2-\frac{ab(ab-c^2)^2}{c^2}\geq 0.
$$
However, the last inequality is not true as it will lead to a
contradiction $ab(ab-c^2)^2\leq c^2(ab-c^2-1)^2 < ab(ab-c^2)^2$.
\hfill$\Box$

The ``only if" parts of statements (i) and (ii) of the next general
result imply respectively that $\mathcal M^{(k)}$ is tightly
monogamous and completely monogamous.

{\bf Theorem C1.} {\it Assume $k\geq 3$. Let
$\rho=\rho_{A_1,A_2,\ldots, A_k}\in{\mathcal {FS}}(H_{A_1}\otimes
H_{A_2}\otimes\cdots\otimes H_{A_k})$ with the CM
$\Gamma_\rho=(A_{ij})_{k\times k}$ as in Eq.(1). For any
$l$-partition  ($2\leq l< k$)   of $k$-partite system $A_1A_2\ldots
A_k$ determined by a permutation $\pi$ of $k$ as in Eq.(6), the
following statements are true.}

 {\rm (i)} {\it With $i_0=0$, we have $$\begin{array}{rl}
 & {\mathcal M}^{(k)}(\rho_{A_1,A_2,\ldots, A_k})= {\mathcal M}^{(l)}(\rho_{A_{\pi(1)}\ldots
A_{\pi(i_1)}|A_{\pi(i_1+1)}\ldots A_{\pi(i_1+i_2)}|\ldots
|A_{\pi(i_1+\cdots +i_{l-1}+1)}\ldots A_{\pi(i_1+\cdots +i_l)}})
\end{array}
$$
 if and only if
$${\mathcal
M}^{(i_h)}(\rho_{A_{\pi(i_0+i_1+\cdots +i_{h-1}+1)}\ldots
A_{\pi(i_0+i_1+\cdots +i_h)}})=0\  {\rm  for \ all}\  h\in\{1,2,\ldots, l\}\ {\rm
with }\ i_h\geq 2.$$}

{\rm (ii)}  {\it For each $h\in\{1, 2,\ldots, l\}$, we have
$${\mathcal M}^{(k)}(\rho_{A_1,A_2,\ldots, A_k})
=  {\mathcal M}^{(i_h)}(\rho_{A_{\pi(i_0+i_1+\cdots
+i_{h-1}+1)}\ldots A_{\pi(i_0+i_1+\cdots +i_h)}})
$$
 if and only
 if, when $h>1$,
$${\mathcal M}^{(2)}(\rho_{A_{\pi(i_1+\cdots +i_{h-1}+1)}\ldots
A_{\pi(i_1+\cdots +i_h)}|A_{\pi(1)}\ldots A_{\pi(i_1+\cdots
+i_{h-1})}A_{\pi(i_1+\cdots +i_{h}+1)}\ldots A_{\pi(k)}})=0 $$ and
$${\mathcal M}^{(k-i_h)}(\rho_{A_{\pi(1)}\ldots
A_{\pi(i_1+\cdots +i_{h-1})}A_{\pi(i_1+\cdots +i_{h}+1)}\ldots
A_{\pi(k)}})=0;$$ when $h=1$,
$${\mathcal M}^{(2)}(\rho_{A_{\pi(1)}\ldots
A_{\pi(i_1)}|A_{\pi(i_1+1)}A_{\pi(i_1+2)}\ldots A_{\pi(k)}})=0\ \ {\rm and}\ \ {\mathcal M}^{(k-i_1)}(\rho_{A_{\pi(i_1+1)}A_{\pi(i_1+2)}\ldots A_{\pi(k)}})=0.$$
}

{\bf Proof.} (i) Denote by $B_h=A_{\pi(i_0+i_1+\cdots
i_{h-1}+1})\cdots A_{\pi(i_0+i_1+\cdots i_{h-1}+i_h)}$ and
$\Gamma_{B_h}$ the CM of $\rho_{B_h}$, $h=1,2,\ldots, l$. Then, by
Eq.(B2),
$$
 {\mathcal M}^{(l)}(\rho_{A_{\pi(1)}\ldots
A_{\pi(i_1)}|A_{\pi(i_1+1)}\ldots A_{\pi(i_1+i_2)}|\ldots
|A_{\pi(i_1+\cdots +i_{l-1}+1)}\ldots A_{\pi(k)}})
 = {\mathcal M}^{(k)}(\rho_{A_1,A_2,\ldots, A_k})
$$
 if and only if $$\Pi_{h=1}^l
\det(\Gamma_{B_h})=\Pi_{h=1}^l(\Pi_{j=i_0+i_1+\cdots+i_{h-1}+1}^{i_1+\cdots+i_{h-1}+i_h}\det(A_{\pi(j),\pi(j)})).$$
By Lemma A1, the above equation holds if and only if
$\det(\Gamma_{B_h})=\Pi_{j=i_0+i_1+\cdots+i_{h-1}+1}^{i_1+\cdots+i_{h-1}+i_h}\det(A_{\pi(j),\pi(j)})$
for each $h=1,2,\ldots, l$, and in turn, if and only if ${\mathcal
M}^{(i_h)}(\rho_{A_{\pi(i_0+i_1+\cdots +i_{h-1}+1)}\ldots
A_{\pi(i_1+\cdots +i_h)}})=0$ for each $h=1,2,\ldots, l$ whenever
$i_h\geq 2$.

(ii) For any $h\in\{1,2,\ldots,l\}$ with $i_h\geq 2$,  denote by
$\Gamma_{B_h^c}=(A_{ij})$ with $i,j\in \{1,2,\ldots, k\}\setminus
B_h=B_h^c$; then
$$\Gamma_\rho\cong\Gamma_{\rho_{B_hB_h^c}}=\left(
\begin{array}{cc} \Gamma_{B_h} & C_h \\ C_h^T & \Gamma_{B_h^c}
\end{array}\right).$$
By Eq.(B3),
$$ {\mathcal
M}^{(i_h)}(\rho_{A_{\pi(i_0+i_1+\cdots +i_{h-1}+1)}\ldots
A_{\pi(i_0+i_1+\cdots +i_h)}}) = {\mathcal M}^{(i_h)}(\rho_{B_h})=
{\mathcal M}^{(k)}(\rho_{A_1A_2\ldots A_k}) \eqno(C1)
$$
if and only if
$$
\det(\Gamma_{B_h^c}-C_h^T\Gamma_{B_h}^{-1}C_h)=\Pi_{j\in B_h^c}
\det(A_{jj}).
$$
By Lemma A1 and Lemma A2, the above equation holds if and only if
$C_h=0$ and $\det(\Gamma_{B_h^c})= \Pi_{j\in B_h^c} \det(A_{jj})$.
Therefore, Eq.(C1) is true if and only if $${\mathcal
M}^{(2)}(\rho_{B_hB_h^c})=0\ {\rm and} \ {\mathcal
M}^{(k-i_h)}(\rho_{B_h^c})=0.$$ This completes the proof of the
statement (ii). \hfill$\Box$

To prove that $\mathcal M^{(k)}$ is not strongly monogamous, the
following general result is useful.

{\bf Theorem C2.} {\it For any $\rho=\rho_{A_1A_2\ldots
A_k}\in{\mathcal {FS}}(H_{A_1}\otimes H_{A_2}\otimes\cdots\otimes
H_{A_k})$ with CM $\Gamma_{\rho}=(A_{ij})_{k\times k}$ as in Eq.(1)
and any $l$-partition ($2\leq l< k$) of $A_1A_2\ldots A_k$
determined by a permutation $\pi$ as in Eq.(6) with $i_l\geq 2$,
 we have
$$\begin{array}{rl}
 & {\mathcal M}^{(l)}(\rho_{A_{\pi(1)}\ldots
A_{\pi(i_1)}|A_{\pi(i_1+1)}\ldots A_{\pi(i_1+i_2)}|\ldots
|A_{\pi(i_1+\cdots +i_{l-1}+1)}\ldots A_{\pi(k)}})\\
= & {\mathcal M}^{(l)}(\rho_{A_{\pi(1)}\ldots
A_{\pi(i_1)}|A_{\pi(i_1+1)}\ldots A_{\pi(i_1+i_2)}|\ldots
|A_{\pi(i_1+\cdots +i_{l-1}+1)}\ldots A_{\pi(k-1)}})
\end{array}
$$
 if and only if
$F=D(\Gamma_{\rho_{A_{\pi(i_1+\cdots +i_{l-1}+1)}\ldots
A_{\pi(k-1)}}})^{-1} E$, where
$$
D=\left(\begin{array}{cccc} A_{\pi(1),\pi(i_1+\cdots +i_{l-1}+1)}& A_{\pi(1),\pi(i_1+\cdots +i_{l-1}+2)} & \cdots & E_{\pi(1),\pi(k-1)} \\
 A_{\pi(2),\pi(i_1+\cdots +i_{l-1}+1)}& A_{\pi(2),\pi(i_1+\cdots +i_{l-1}+2)} & \cdots & E_{\pi(2),\pi(k-1)} \\
\vdots & \vdots & \ddots & \vdots \\
A_{\pi(i_1+\cdots +i_{l-1}),\pi(i_1+\cdots +i_{l-1}+1)}&
A_{\pi(i_1+\cdots +i_{l-1}),\pi(i_1+\cdots +i_{l-1}+2)} & \cdots &
A_{\pi(i_1+\cdots +i_{l-1}),\pi(k-1)}
\end{array}\right) $$
is an $(k-i_l)\times (i_l-1)$ block matrix,
$$ F=\left(\begin{array}{c} A_{\pi(1),\pi(k)} \\ A_{\pi(2),\pi(k)} \\ \vdots \\
A_{\pi(i_1+\cdots +i_{l-1}),\pi(k)}
\end{array}\right) \quad{\rm and}\quad E=\left(\begin{array}{c} A_{\pi(i_1+\cdots +i_{l-1}+1),\pi(k)} \\ A_{\pi(i_1+\cdots +i_{l-1}+2),\pi(k)} \\
\vdots \\ A_{\pi(k-1),\pi(k)}
\end{array}\right).
$$
}

The following result is a special case of Theorem C1, which
illustrates  the exact meaning of Theorem C2 plainly.

{\bf Theorem C3.} {\it For $k\geq 2$ and $\rho=\rho_{A_1A_2\ldots
A_kA_{k+1}}\in{\mathcal {FS}}(H_{A_1}\otimes H_{A_2}\otimes
\cdots\otimes H_{A_k}\otimes H_{A_{k+1}})$ with CM
$\Gamma_\rho=(A_{ij})_{(k+1)\times (k+1)}$,
$$
{\mathcal M}^{(k)}(\rho_{A_1A_2\ldots A_{k-1}|A_kA_{k+1}})=
{\mathcal M}^{(k)}(\rho_{A_1A_2\ldots A_k})
\  \mbox{ if and only if
}\
\left(\begin{array}{c} A_{1,k+1} \\ A_{2,k+1} \\ \vdots \\
A_{k-1,k+1}
\end{array}\right) =\left(\begin{array}{c} A_{1,k} \\ A_{2,k} \\ \vdots \\
A_{k-1,k}
\end{array}\right)  A_{kk}^{-1}A_{k,k+1}.
$$}

{\bf Proof of Theorem C1.} Without loss of generality, we may assume
that the $l$-partition of $A_1A_2\ldots A_{k-1}A_k$ is
$$A_1\ldots
A_{i_1}|A_{i_1+1}\ldots A_{i_1+i_2}|A_{i_1+i_2+1}\ldots
A_{i_1+\cdots+i_{l-1}}|A_{i_1+\cdots+i_{l-1}+1}\ldots
A_{k-1,k-1}A_{kk}$$ since $\mathcal M^{(k)}$ is invariant under any
permutation of subsystems. It is clear by the assumption $i_l\geq2$
that $i_1+i_2+\cdots+i_{l-1}<k-1$. With the same symbols used in the
proof of Theorem B5, and by Eqs.(B5)-(B8),
$$\begin{array}{rl}
& {\mathcal M}^{(l)}(\rho_{A_1\ldots A_{i_1}|A_{i_1+1}\ldots
A_{i_1+i_2}|A_{i_1+i_2+1}\ldots
A_{i_1+\cdots+i_{l-1}}|A_{i_1+\cdots+i_{l-1}+1}\ldots
A_{k-1,k-1}A_{kk}})\\ = & {\mathcal M}^{(l)}(\rho_{A_1\ldots
A_{i_1}|A_{i_1+1}\ldots A_{i_1+i_2}|A_{i_1+i_2+1}\ldots
A_{i_1+\cdots+i_{l-1}}|A_{i_1+\cdots+i_{l-1}+1}\ldots A_{k-1,k-1}})
\end{array}
$$
 if and only if
$$
\frac{\det(\Lambda_{\rho})}{
\det(\Lambda_{\rho_{B_l}})}=\frac{\det(\Gamma_{\rho})}{
\det(\Gamma_{B_l})}=
\frac{\det(\Gamma_{\rho_{k^c}})}{\det(\Gamma_{\rho_{A_{i_1+\cdots
+i_{l-1}+1}\ldots
A_{k-1£¬k-1}}})}=\frac{\det(\Lambda_{\rho_{k^c}})}{\det(\Lambda_{\rho_{A_{i_1+\cdots
+i_{l-1}+1}\ldots A_{k-1,k-1}}})}, \eqno(C2)
$$
where
$$
\Lambda_\rho=\left(\begin{array}{cccc} I & E_{12} & \cdots &
E_{1k} \\ E_{12}^T & I & \cdots & E_{2k} \\
\vdots & \vdots & \ddots & \vdots \\
E_{1k}^T & E_{2k}^T & \cdots & I
\end{array}\right)=\left(\begin{array}{ccc} \Lambda_{\rho_{A_1\ldots A_{i_1+\cdots+i_{l-1}}}}& X & Z \\
X^T & \Lambda_{\rho_{A_{i_1+\cdots +i_{l-1}+1}\ldots A_{k-1}}} &  Y \\
Z^T & Y^T &   I
\end{array}\right)
$$
with $E_{ij}=A_{ii}^{-\frac{1}{2}}A_{ij}A_{jj}^{-\frac{1}{2}}$,
$i\not=j$,
$$
X=\left(\begin{array}{cccc} E_{1,i_1+\cdots +i_{l-1}+1}& E_{1,i_1+\cdots +i_{l-1}+2} & \cdots & E_{1,k-1} \\
E_{2,i_1+\cdots +i_{l-1}+1}& E_{2,i_1+\cdots +i_{l-1}+2} & \cdots & E_{2,k-1} \\
\vdots & \vdots & \ddots & \vdots \\
E_{i_1+\cdots +i_{l-1},i_1+\cdots +i_{l-1}+1}& E_{i_1+\cdots
+i_{l-1},i_1+\cdots +i_{l-1}+2} & \cdots & E_{i_1+\cdots
+i_{l-1},k-1}
\end{array}\right), $$
$$ Z=\left(\begin{array}{c} E_{1k} \\ E_{2k} \\ \vdots \\
E_{i_1+\cdots +i_{l-1},k}
\end{array}\right) \quad{\rm and}\quad Y=\left(\begin{array}{c} E_{i_1+\cdots +i_{l-1}+1,k} \\ E_{i_1+\cdots +i_{l-1}+2,k} \\ \vdots \\ E_{k-1,k}
\end{array}\right).
$$
It is clear that Eq.(C2) holds if and only if
$$
\left(\begin{array}{cc} Z^T& Y^T
\end{array}\right)
\Lambda_{\rho_{k^c}}^{-1}\left(\begin{array}{c} Z \\ Y
\end{array}\right) = Y^T
\Lambda_{\rho_{A_{i_1+\cdots +i_{l-1}+1}\ldots A_{k-1,k-1}}}^{-1} Y.
\eqno(C3)
$$
By Lemma B2, Eq.(C3) is true if and only if
$Z=X\Lambda_{\rho_{A_{i_1+\cdots +i_{l-1}+1}\ldots
A_{k-1,k-1}}}^{-1} Y$, which is equivalent to
$$
F=D(\Gamma_{\rho_{A_{i_1+\cdots +i_{l-1}+1}\ldots
A_{k-1,k-1}}})^{-1}E, \eqno(C4)
$$
where $$
D=\left(\begin{array}{cccc} A_{1,i_1+\cdots +i_{l-1}+1}& A_{1,i_1+\cdots +i_{l-1}+2} & \cdots & A_{1,k-1} \\
A_{2,i_1+\cdots +i_{l-1}+1}& A_{2,i_1+\cdots +i_{l-1}+2} & \cdots & A_{2,k-1} \\
\vdots & \vdots & \ddots & \vdots \\
A_{i_1+\cdots +i_{l-1},i_1+\cdots +i_{l-1}+1}& A_{i_1+\cdots
+i_{l-1},i_1+\cdots +i_{l-1}+2} & \cdots & A_{i_1+\cdots
+i_{l-1},k-1}
\end{array}\right), $$
$$ F=\left(\begin{array}{c} A_{1k} \\ A_{2k} \\ \vdots \\
A_{i_1+\cdots +i_{l-1},k}
\end{array}\right) \quad{\rm and}\quad E=\left(\begin{array}{c} A_{i_1+\cdots +i_{l-1}+1,k} \\ A_{i_1+\cdots +i_{l-1}+2,k} \\ \vdots \\ A_{k-1,k}
\end{array}\right).
$$
Hence the statement of the theorem is true and the proof is
completed.\hfill$\Box$

Now we are in a position to prove Theorem 3.

{\bf Proof of Theorem 3.} Assume $k\geq 3$. It is obvious that
$\mathcal M^{(k)}$ is tightly monogamous by  (i) and is completely
monogamous by  (ii) in Theorem C1.  Hence, the statements (1) and (2)
of Theorem 3 are true.

To prove the statement (3), assume $k\geq 2$. Let
$\rho_{A_1A_2\ldots A_kA_{k+1}}\in{\mathcal {FS}}(H_{A_1}\otimes
H_{A_2}\otimes\cdots\otimes H_{A_k}\otimes H_{A_{k+1}})$ with CM
$\Gamma_\rho=(A_{ij})_{(k+1)\times (k+1)}$. It is clear that
$\mathcal M^{(2)}(\rho_{A_1A_{k+1}})= \mathcal
M^{(2)}(\rho_{A_2A_{k+1}})=\cdots=\mathcal
M^{(2)}(\rho_{A_kA_{k+1}})=0$ if and only if
$$
\left(\begin{array}{c} A_{1,k+1} \\ A_{2,k+1} \\ \vdots \\
A_{k-1,k+1}
\end{array}\right) =\left(\begin{array}{c} 0 \\ 0 \\ \vdots \\
0
\end{array}\right)=0.
$$

However, by Theorem C3, $ {\mathcal M}^{(k)}(\rho_{A_1A_2\ldots
A_{k-1}|A_kA_{k+1}})= {\mathcal M}^{(k)}(\rho_{A_1A_2\ldots A_k}) $
 if and only if
$$
\left(\begin{array}{c} A_{1,k+1} \\ A_{2,k+1} \\ \vdots \\
A_{k-1,k+1}
\end{array}\right) =\left(\begin{array}{c} A_{1,k}  A_{kk}^{-1}A_{k,k+1} \\ A_{2,k}  A_{kk}^{-1}A_{k,k+1} \\ \vdots \\
A_{k-1,k}  A_{kk}^{-1}A_{k,k+1}
\end{array}\right),
$$
which may not be zero. To make this sure, let us give an example
here. Consider the $(k+1)$-partite $(1+1+\cdots+1)$-mode case. Let
$\Gamma=(A_{ij})$ be a real symmetric matrix, where,
$A_{jj}=\left(\begin{array}{cc} 2 &0
\\ 0 & 2 \end{array}\right)$ for $j=1,2,\ldots, k-1,k+1$, $A_{kk}=\left(\begin{array}{cc} 3 &0
\\ 0 & 3 \end{array}\right)$, $A_{k-1,k}=A_{k,k+1}=\left(\begin{array}{cc} 1 &0
\\ 0 & 1 \end{array}\right)$, $A_{k-1,k+1}=\left(\begin{array}{cc} \frac{1}{3} &0
\\ 0 & \frac{1}{3} \end{array}\right)$, otherwise, $A_{ij}=0$ for
  $i<j$. It is easily checked that $\Gamma=\Gamma_\rho$ is a CM for a Gaussian state
  $\rho=\rho_{A_1A_2,\ldots,A_kA_{k+1}}$ since $\Gamma+i \Delta\geq 0$, where $\Delta=\oplus_{j=1}^{k+1}\Delta_j$ with $\Delta_j=\left(\begin{array}{cc}
  0&1
\\ -1 & 0 \end{array}\right)$. Obviously, we have
$$
\left(\begin{array}{c} A_{1,k+1} \\ A_{2,k+1} \\ \vdots \\
A_{k-1,k+1}
\end{array}\right) =\left(\begin{array}{c} A_{1,k}  A_{kk}^{-1}A_{k,k+1} \\ A_{2,k}  A_{kk}^{-1}A_{k,k+1} \\ \vdots \\
A_{k-1,k}  A_{kk}^{-1}A_{k,k+1}
\end{array}\right)\not=0
$$
as $A_{k-1,k+1}=A_{k-1,k}  A_{kk}^{-1}A_{k,k+1}\not=0$. In fact, for
this state $\rho$, we have ${\mathcal M}^{(k)}(\rho_{A_1A_2\ldots
A_{k-1}|A_kA_{k+1}})={\mathcal M}^{(k)}(\rho_{A_1A_2\ldots
A_{k-1}A_k})\approx 0.3056$, but $\mathcal
M^{(2)}(\rho_{A_{k-1}A_{k+1}})\approx 0.0548\not=0$.

Hence $\mathcal M^{(k)}$ is not strongly monogamous, which completes
the proof of the statement (3) in Theorem 3.
 \hfill$\Box$

\end{widetext}

\end{document}